\documentclass[aps,pra,reprint,superscriptaddress,amsmath,amssymb]{revtex4-2}
\usepackage[T1]{fontenc}
\usepackage{lmodern}
\usepackage[utf8]{inputenc}
\usepackage{graphicx,booktabs,calc}
\usepackage[unicode,hidelinks]{hyperref}
\begin{document}
\title{Quantum CSS codes: Decoder-aware multiobjective design and bound-certified selection}
\author{Jiapei Zhuang}
\thanks{J. Z. and Y. C. contributed equally to this work.\\ Corresponding author: \href{mailto:yaochengchen@usp.br}{yaochengchen@usp.br}\\ Corresponding author: \href{mailto:jpzhuangphy@gmail.com}{jpzhuangphy@gmail.com}}
\affiliation{Quantum Science Center of Guangdong-Hong Kong-Macao Greater Bay Area, Shenzhen 518045, China}
\author{Yaocheng Chen}
\thanks{J. Z. and Y. C. contributed equally to this work.\\ Corresponding author: \href{mailto:yaochengchen@usp.br}{yaochengchen@usp.br}\\ Corresponding author: \href{mailto:jpzhuangphy@gmail.com}{jpzhuangphy@gmail.com}}
\affiliation{Instituto de Astronomia, Geofísica e Ciências Atmosféricas (IAG-USP), Universidade de São Paulo, Rua do Matão 1226, São Paulo, SP 05508-090, Brazil}
\author{Xingyao Wu}
\email[Corresponding author: ]{xingyao.wu@sdu.edu.cn}
\affiliation{School of Computer Science and Technology, Shandong University, Qingdao 266237, China}
\begin{abstract}
Quantum error-correcting codes trade decoding reliability against the resources needed to measure their checks. We formulate Calderbank--Shor--Steane (CSS) code design at fixed encoding resources as a decoder-aware multiobjective problem. The objectives are failure probability under sector-wise degenerate maximum-likelihood decoding and check weight, the minimum total weight of an independent pure-type generator basis. For independent bit- and phase-flip noise with perfect syndrome extraction, deterministic truncation bounds support selection under a failure-rate constraint. At a per-sector error probability of 0.01, the bounds certify that a code encoding two logical qubits into 40 physical qubits meets a failure-rate target of 0.001 at check weight 139. The lower bound of every cheaper candidate in the corresponding \((40,2)\) main menu exceeds this target. In paired searches sharing one annealing kernel and equal query budgets, decoder-aware guidance gives significantly higher mean final hypervolume than distance guidance in all eight tested configurations under the reported Holm correction. Against distance--multiplicity guidance, five configurations show significant gains and three do not, including one with a negative mean difference. All arms are scored in common truncated-reliability and normalized-check-weight coordinates. The framework connects reliability--check-weight trade-offs to bound-certified selection within evaluated CSS pools.
\end{abstract}
\maketitle
\section{Introduction}\label{i.-introduction}

Quantum error correction protects logical information through redundant encoding and recovery from physical errors \cite{terhal2015}. Recent demonstrations of below-threshold surface-code memories \cite{google2025}, error-corrected logical processing in neutral-atom arrays \cite{bluvstein2026}, and concatenated bosonic memories \cite{putterman2025} make the resource cost of reliable quantum information processing an increasingly concrete design problem. That cost extends beyond the number of physical qubits. Recent work addresses encoding overhead and hardware connectivity through quantum low-density parity-check codes \cite{bravyi2024,pecorari2025,scruby2026}, the number of syndrome-extraction rounds through transversal algorithmic fault tolerance \cite{zhou2025}, and classical processing through parallel decoding \cite{hillmann2025}. Across these directions, assessing a design requires specifying both the reliability sought and the resources counted, together with the noise model, measurement procedure, and decoder.

Automated code design offers ways to explore these choices, but existing objectives capture different aspects of performance. Several lines of work show how structural or error-sensitive information enters automated design. Distance-based annealing pairs a multiplicity surrogate with generator reduction \cite{perlin}. Evolutionary search optimizes distance together with noise-tailored undetectable errors \cite{webster}. Reinforcement learning combines degree reduction with distance rewards \cite{he}. High-rate constructions also explicitly consider parameter trade-offs and the retention of Pareto sets \cite{lee}. Decoder feedback has been used to adapt surface-code memories toward a target logical error rate \cite{nautrup}, while quantum Lego searches optimize distance and biased-noise objectives \cite{su}.

Comparing these approaches requires attention to their reliability metrics. In the biased-noise task of Su et al., the reward uses logical-error probability conditioned on a trivial syndrome; the work also reports undetectable logical-error probability [Eqs. (1)--(3) of Ref. \cite{su}]. These quantities differ from the full degenerate maximum-likelihood (DML) failure probability over all syndromes used here. Their framework also accommodates decoder-based evaluations including nontrivial syndromes. We therefore connect a specified reliability objective, check cost, and bound-based selection within a single design framework.

At fixed Calderbank--Shor--Steane (CSS) encoding resources, how should one choose between logical reliability and the cost of measuring checks? Code distance alone does not determine a decoder's failure probability under a specified noise model. Minimizing failure alone, in turn, does not express how much measurement cost is justified by a user's reliability requirement. A fixed preference weight selects one compromise, whereas different budgets or failure requirements can favor different codes. Keeping reliability and check cost as separate objectives exposes these choices and permits selection after the requirement is specified. A constrained single-objective problem can address a particular requirement; the multiobjective formulation retains alternatives across requirements.

Three elements are needed to make such choices meaningful: a cost definition invariant under equivalent generator representations, reliability evaluated for a specified decoder, and decision rules that account for evaluation uncertainty. Changing a stabilizer basis can alter its total generator weight without changing the code. Likewise, a truncated failure estimate can leave threshold attainment unresolved. We connect these elements by minimizing \((P_D,B_*)\) at fixed block length, encoded dimension, and stabilizer ranks. Here \(P_D\) is the block failure probability of sector-wise DML decoding under independent bit- and phase-flip noise with perfect syndrome extraction, and \(B_*\) is the minimum total weight of an independent pure-type stabilizer basis. A distance floor and exclusion of components with no logical information define the feasible family. Low-weight error enumeration supplies two-sided bounds for threshold selection and exclusion of candidates dominated under the interval test. These bounds control analytic truncation error; their numerical scope and the interpretation of the dominance test are specified in Sec.~\ref{iii.-certified-evaluation-and-decision-rules}.

Our contributions are threefold. First, we formulate fixed-resource CSS design with a decoder-defined reliability objective and a basis-invariant check-weight objective, subject to distance and component constraints. Second, we derive two-sided truncation bounds and use them for threshold and dominance decisions. Third, we compare decoder-aware guidance with alternative signals in paired, equal-budget searches and test joint code--redundancy selection within a fixed evaluated pool.

The resulting menus (retained candidate sets) support a concrete design decision: at \((n,k)=(40,2)\) and \(p=0.01\) per sector, a check-weight-139 code meets a \(10^{-3}\) failure requirement and every cheaper main-menu candidate is excluded by its lower bound. In paired common-kernel searches, decoder-aware guidance raises mean final application hypervolume relative to distance in all eight tested cells under the reported Holm correction; relative to distance--multiplicity, five cells show significant gains and three do not, including a negative mean difference at \((40,4)\). The comparison uses common truncated reliability coordinates, while the selection uses refined failure bounds.

Noisy syndrome measurements introduce another resource decision: how much of a measured-check-weight budget to allocate to redundancy. Building on quantum data-syndrome coding \cite{ashikhmin}, we re-evaluate the 477 retained candidates from the eight main menus under a strict single-round model and compare joint code-and-redundancy selection with selecting the code first. Interval-separated comparisons support gains at many tested budgets, while other budgets show no endpoint gain.  Together, the analyses connect decoder-defined reliability to explicit check-weight decisions under the stated models.

Section~\ref{ii.-decoder-aware-multiobjective-formulation} defines the multiobjective problem. Section~\ref{iii.-certified-evaluation-and-decision-rules} develops the evaluation bounds and decision rules, and Sec.~\ref{iv.-search-and-evaluation-protocol} specifies the search comparison. Sections~\ref{v.-results} and~\ref{vi.-extension-to-noisy-syndrome-measurements} present the code-capacity (perfect-syndrome) results and measurement-noise extension. Sections~\ref{vii.-discussion} and~\ref{viii.-conclusion} discuss the implications and conclude. Appendixes~\ref{app:search-execution}--\ref{app:noise-summaries} give search details, sensitivity results, and measurement-noise summaries.

\section{Decoder-aware multiobjective
formulation}\label{ii.-decoder-aware-multiobjective-formulation}

We formulate CSS code design as a choice between logical reliability and
check weight at fixed encoding resources. The reliability objective is
defined by a specified noise model and decoder; the cost objective is
minimized over equivalent generator bases. Keeping these objectives separate accommodates different resource budgets without fixing a single trade-off weight in advance.

\subsection{Design variables and noise
model}\label{a.-design-variables-and-noise-model}

In the CSS construction \cite{calderbank1996,steane1996}, a code is specified by two binary subspaces, \(C=(S_X,S_Z)\),
represented by generator matrices \(H_X\) and \(H_Z\) satisfying

\begin{equation}
\begin{gathered}
S_X=\operatorname{row}(H_X),\\
S_Z=\operatorname{row}(H_Z),\\
H_XH_Z^{T}=0.
\end{gathered}
\label{eq:v9-01}
\end{equation}

The subspaces specify the supports of the \(X\)-type and \(Z\)-type
stabilizers. Their dimensions are \(r_X\) and \(r_Z\), and the code
encodes \(k=n-r_X-r_Z\) logical qubits in \(n\) physical qubits. We fix
\((n,k,r_X,r_Z)\) within each design problem. The numerical study uses
balanced ranks, \(r_X=r_Z=(n-k)/2\), which requires \(n-k\) to be even; every tested cell meets this condition. Invertible row operations change
the representation of a code but leave its stabilizer subspaces unchanged \cite{gottesman1997}.

We consider independent \(X\) and \(Z\) Bernoulli errors, each with
probability \(p\) per physical qubit, and perfect syndrome extraction.
Unless otherwise stated, \(p=0.01\). We call the \(X\)- and \(Z\)-type error sets sectors. The decoder performs degenerate
maximum-likelihood (DML) decoding separately in each sector. For
each syndrome, it selects the error class modulo the corresponding
stabilizer subspace with the largest total probability. This rule accounts for degeneracy by maximizing the probability of an equivalence class rather than that of an individual error \cite{iyer2015}. Thus the
reliability objective depends on both the code and the specified
decoding problem.

We regard binary vectors as coordinate tuples; they are written as
columns in matrix--vector products, while stabilizer subspaces are
specified by row spaces. For an \(X\) error \(e\in\mathbb F_2^n\), let
\(\sigma(e)=H_Ze\) denote its syndrome. Two errors with the same syndrome belong to the same class if their difference lies in \(S_X\). We reserve \(s\) for a syndrome, \(\ell\) for a
stabilizer-class label, and \(\alpha\in\{X,Z\}\) for a sector index; \([e]\) denotes the stabilizer class containing \(e\). Define the class probability masses by

\begin{equation}
\begin{gathered}
M(s,\ell)=
\sum_{\substack{e:\,\sigma(e)=s\\{}[e]=\ell}}
p^{|e|}(1-p)^{n-|e|},
\\
P^X=1-\sum_s\max_\ell M(s,\ell),
\end{gathered}
\label{eq:v9-02}
\end{equation}

where \(|e|\) is the Hamming weight. The \(Z\)-sector failure
probability \(P^Z\) is defined by interchanging \(X\) and \(Z\).
Independence of the two error sectors gives the block failure
probability

\begin{equation}
\begin{gathered}
P_D(C)=1-[1-P^X(C)][1-P^Z(C)].
\end{gathered}
\label{eq:v9-03}
\end{equation}

The subscript \(D\) refers throughout to this sector-wise DML decoder.
These definitions specify the exact reliability objective; its
computational evaluation is addressed in the next section.

\subsection{Reliability and basis-invariant check
cost}\label{b.-reliability-and-basis-invariant-check-cost}

Our second objective is the minimum total weight of independent
pure-type stabilizer generators,

\begin{equation}
\begin{gathered}
B_*(C)=
\min_{\mathcal B_X\text{ basis of }S_X}
\sum_{g\in\mathcal B_X}|g|
+
\min_{\mathcal B_Z\text{ basis of }S_Z}
\sum_{g\in\mathcal B_Z}|g|.
\end{gathered}
\label{eq:v9-04}
\end{equation}

This minimization removes dependence on the particular matrices used to represent the code. For each subspace, inserting linearly independent
nonzero elements in nondecreasing Hamming weight yields a minimum-weight
basis by the vector-matroid greedy rule \cite{oxley}. The two
minimizations can be carried out independently. Our implementation enumerates the nonzero
elements of each stabilizer subspace and applies this rule; its
enumeration cost is exponential in the subspace rank.

\(B_*\) counts data-qubit participations in one round of independent
pure-type checks, with each participation assigned unit weight. It
provides a common cost proxy for comparing codes at fixed ranks.
Routing, ancilla requirements, measurement scheduling, and error
propagation depend on the syndrome-extraction circuit and are not
included in this objective.  Redundant checks introduce an additional
choice of measured generators, considered later in the measurement-noise
extension.

For a feasible set \(\mathcal F\) defined below, the design problem is

\begin{equation}
\begin{gathered}
\min_{C\in\mathcal F}\bigl(P_D(C),B_*(C)\bigr).
\end{gathered}
\label{eq:v9-05}
\end{equation}

Both objectives are properties of the stabilizer subspaces under the
stated noise and decoding assumptions. Treating them separately retains
alternatives with different reliability and check-weight requirements.
Search generates candidate codes for this multiobjective problem; the objectives are defined independently of the search method.

\subsection{Distance and component
constraints}\label{c.-distance-and-component-constraints}

We impose a minimum distance \(d(C)\ge3\). Distance specifies a protection requirement, while \(P_D\) resolves reliability differences within the
feasible set. We also exclude factors that consume physical qubits
without encoding logical information. A frozen qubit, for example,
contributes check weight but no logical degree of freedom. Including
such factors would allow part of a nominally fixed block length to
remain unused for encoding.

To define this restriction independently of a generator basis, consider
a partition \(\mathcal P\) of the physical qubits. The code decomposes
over \(\mathcal P\) when

\begin{equation}
\begin{gathered}
S_X=\bigoplus_{B\in\mathcal P}
\left(S_X\cap\mathbb F_2^B\right),\\
S_Z=\bigoplus_{B\in\mathcal P}
\left(S_Z\cap\mathbb F_2^B\right),
\end{gathered}
\label{eq:v9-06}
\end{equation}

where \(\mathbb F_2^B\) denotes vectors supported on \(B\). The
components of \(C\) are the blocks of the finest such partition. A
component \(c\) with \(n_c\) physical qubits and restricted stabilizer
ranks \(r_{X,c}\) and \(r_{Z,c}\) encodes

\begin{equation}
\begin{gathered}
k_c=n_c-r_{X,c}-r_{Z,c}
\end{gathered}
\label{eq:v9-07}
\end{equation}

logical qubits. The feasible set consists of CSS codes with the fixed
ranks above satisfying

\begin{equation}
\begin{gathered}
\mathcal F=
\left\{C:\ d(C)\ge3,\quad k_c\ge1
\text{ for every component }c\right\}.
\end{gathered}
\label{eq:v9-08}
\end{equation}

This condition excludes frozen qubits and other stabilizer-state factors. It still admits direct sums whose components each encode logical information. For each stabilizer subspace, its finest coordinate
decomposition is the component partition of the column matroid
represented by a basis matrix. The finest partition decomposing both
subspaces is the join of these two partitions, where join means the
finest common coarsening. Appendix~\ref{appendix-a.-basis-invariant-coordinate-components} states and proves this result and
gives a construction from reduced row-echelon bases. The connection to
fundamental circuits uses standard matroid terminology \cite{oxley}; the
CSS decomposition claim is proved explicitly in the appendix. 

\subsection{Pareto comparison and resource-constrained
choice}\label{d.-pareto-comparison-and-resource-constrained-choice}

A feasible code \(C'\) dominates \(C\) if

\begin{equation}
\begin{gathered}
P_D(C')\le P_D(C),\qquad B_*(C')\le B_*(C),
\end{gathered}
\label{eq:v9-09}
\end{equation}

with at least one strict inequality. Pareto-optimal codes are those not
dominated by another member of the set under consideration. We
distinguish the full feasible set \(\mathcal F\) from a finite evaluated
candidate set \(\mathcal A\subseteq\mathcal F\). After interval filtering, we call the retained candidates in a specified evaluated candidate set a menu. The numerical search
explores only part of \(\mathcal F\), so comparisons among its
candidates are relative to \(\mathcal A\).

When exact objective values are available, a check-weight budget \(b\)
selects a code through

\begin{equation}
\begin{gathered}
C_b\in\underset{C\in\mathcal A:\,B_*(C)\le b}
{\operatorname{arg\,min}}\ P_D(C),
\end{gathered}
\label{eq:v9-10}
\end{equation}

with ties resolved by smaller \(B_*\). Conversely, a maximum allowed
failure probability \(\tau\) selects

\begin{equation}
\begin{gathered}
C_\tau\in\underset{C\in\mathcal A:\,P_D(C)\le\tau}
{\operatorname{arg\,min}}\ B_*(C),
\end{gathered}
\label{eq:v9-11}
\end{equation}

with ties resolved by smaller \(P_D\). Whenever the corresponding
constraint set is nonempty, either rule selects a Pareto-optimal member
of \(\mathcal A\): a dominating code would either improve the minimized
objective or win the tie break.

\begin{figure*}[t]
\centering
\includegraphics[width=\linewidth]{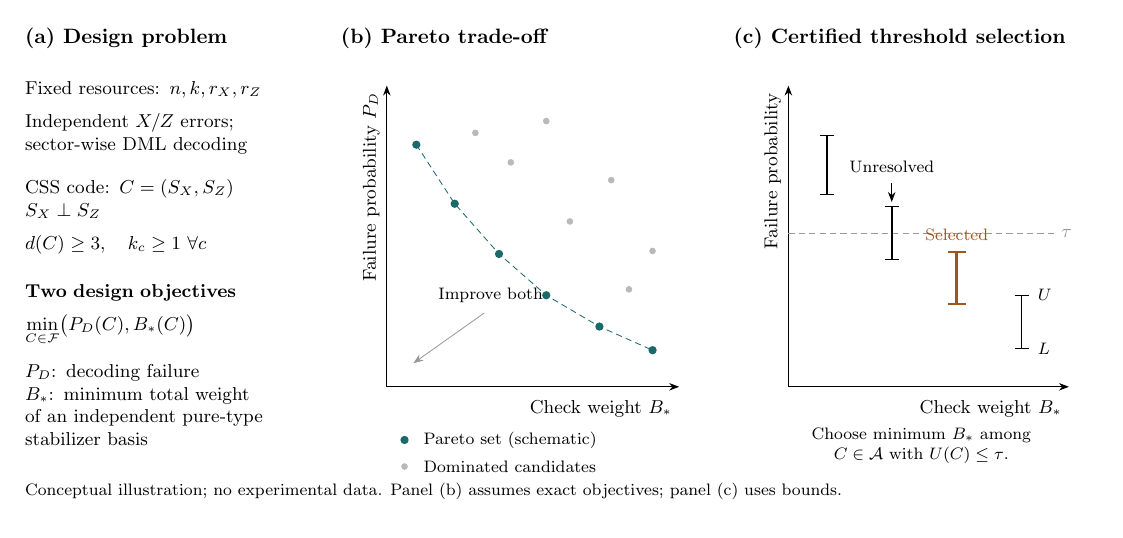}
\caption{Decoder-aware design and conservative selection. (a) Fixed encoding resources, independent \(X/Z\) noise with perfect syndrome extraction, and sector-wise DML decoding define the task; reliability and basis-invariant check weight are separate objectives. (b) Exact-objective schematic for a finite candidate set: green points are nondominated, gray points are dominated, and both objectives improve toward the lower left. The dashed line is a guide, not an interpolation between realizable codes. (c) Illustrative selection with failure bounds: the cheapest candidate fails the requirement, the next is unresolved, and the highlighted point minimizes cost among certified-feasible choices. The unresolved cheaper candidate prevents a minimum-cost conclusion among all truly feasible candidates. All coordinates are illustrative.}
\label{fig:formulation}
\end{figure*}

Figure~\ref{fig:formulation} connects the two-objective design problem to finite-pool selection. Panel (b) illustrates nondominance when the objectives are known exactly. In practice, evaluating \(P_D\) exactly becomes costly as \(n\) increases, so the next section develops deterministic bounds for conservative decisions. Panel (c) shows why meeting an upper-bound requirement certifies feasibility, whereas an unresolved cheaper candidate can prevent a minimum-cost conclusion over all truly feasible candidates. The bounds determine which claims about the true \(P_D\) objective the available evaluations support.

\section{Certified evaluation and decision
rules}\label{iii.-certified-evaluation-and-decision-rules}

The formulation in Sec.~\ref{ii.-decoder-aware-multiobjective-formulation} requires the failure probability of the full
DML decoder. We evaluate this objective through a truncated error
distribution with a deterministic remainder bound. The resulting
intervals support two distinct operations: certifying that a code meets
a reliability threshold and excluding candidates dominated under the interval test. 

\subsection{Truncated sector
probabilities}\label{a.-truncated-sector-probabilities}

For one error sector, let \(M(s,\ell)\) be the syndrome--class
probability mass defined in Sec.~\ref{ii.-decoder-aware-multiobjective-formulation}. Retaining errors of weight at most
\(W\), with \(0\le W\le n\), gives

\begin{equation}
\begin{gathered}
M_W(s,\ell)=\sum_{\substack{e:\,\sigma(e)=s,\ [e]=\ell\\ |e|\le W}}
p^{|e|}(1-p)^{n-|e|}.
\end{gathered}
\label{eq:v9-12}
\end{equation}

The retained probability and omitted tail are

\begin{equation}
\begin{gathered}
m_W=\sum_{w=0}^{W}\binom nw p^w(1-p)^{n-w},
\\ \tau_W=1-m_W.
\end{gathered}
\label{eq:v9-13}
\end{equation}

We use the unnormalized retained distribution. With

\begin{equation}
\begin{gathered}
A_W=\sum_s\max_\ell M_W(s,\ell),\\
P_W=m_W-A_W,
\end{gathered}
\label{eq:v9-14}
\end{equation}

\(P_W\) is the failure mass of the ML rule optimized for the retained distribution. Because omitted errors can change the maximizing class, \(P_W\) is neither the conditional failure probability given weight at most \(W\) nor generally equal to the low-weight contribution to full DML failure.

\textbf{Proposition 1 (sector bracket).} For the full DML failure
\(P=1-\sum_s\max_\ell M(s,\ell)\),

\begin{equation}
\begin{gathered}
P_W\le P\le P_W+\tau_W.
\end{gathered}
\label{eq:v9-15}
\end{equation}

The lower endpoint is nondecreasing in \(W\), the upper endpoint is
nonincreasing, and both equal \(P\) at \(W=n\).

\emph{Proof.} Set \(A=\sum_s\max_\ell M(s,\ell)\). For each syndrome,

\begin{equation}
\begin{gathered}
0\le\max_\ell M(s,\ell)-\max_\ell M_W(s,\ell)
\\ \le\sum_\ell\bigl[M(s,\ell)-M_W(s,\ell)\bigr].
\end{gathered}
\label{eq:v9-16}
\end{equation}

Summing gives \(0\le\delta_W:=A-A_W\le\tau_W\). Therefore

\begin{equation}
\begin{gathered}
P=1-A=P_W+\tau_W-\delta_W,
\end{gathered}
\label{eq:v9-17}
\end{equation}

which proves the bracket. If the cutoff increases from \(W\) to \(W'\),
the additional retained mass is \(\Delta m=m_{W'}-m_W\ge0\), and
\(0\le A_{W'}-A_W\le\Delta m\). Hence \(P_{W'}-P_W\ge0\). The upper
endpoint satisfies \(P_W+\tau_W=1-A_W\) and therefore cannot increase.
At \(W=n\) the omitted tail vanishes. \(\square\)

\subsection{Combining the two
sectors}\label{b.-combining-the-two-sectors}

For the independent \(X\) and \(Z\) sectors of Sec.~\ref{ii.-decoder-aware-multiobjective-formulation}, define

\begin{equation}
\begin{gathered}
L_W=\hat P_W=1-(1-P_W^X)(1-P_W^Z),
\end{gathered}
\label{eq:v9-18}
\end{equation}

\begin{equation}
\begin{gathered}
U_W=1-(1-P_W^X-\tau_W)(1-P_W^Z-\tau_W).
\end{gathered}
\label{eq:v9-19}
\end{equation}

Each sector upper endpoint lies in \([0,1]\), since
\(P_W^\alpha+\tau_W=1-A_W^\alpha\) for \(\alpha=X,Z\). The function
\(1-(1-x)(1-z)\) is nondecreasing in both arguments on this domain, so
Proposition 1 gives

\begin{equation}
\begin{gathered}
L_W\le P_D\le U_W.
\end{gathered}
\label{eq:v9-20}
\end{equation}

In particular,

\begin{equation}
\begin{gathered}
U_W-L_W
=\tau_W(2-P_W^X-P_W^Z)-\tau_W^2
\le2\tau_W.
\end{gathered}
\label{eq:v9-21}
\end{equation}

Thus \(\min\{1,L_W+2\tau_W\}\) is a valid, generally looser block upper bound: both sector tails contribute to block failure, so a single-sector tail \(\tau_W\) cannot generally serve as its additive bound. The block intervals are nested as \(W\)
increases, because the sector lower endpoints increase and their upper
endpoints decrease. At \(W=n\), \(L_n=U_n=P_D\).

These bounds quantify omitted probability mass under the independent-sector model of Sec.~\ref{ii.-decoder-aware-multiobjective-formulation}.

\subsection{Evaluation and numerical
checks}\label{c.-evaluation-and-numerical-checks}

The truncated evaluator groups error patterns by syndrome and stabilizer
class. For rational \(p=a_0/b_0\) with integers \(0<a_0<b_0\), class
comparisons can be performed using the integer masses

\begin{equation}
\begin{gathered}
\widetilde M_W(s,\ell)
=\sum_{w=0}^{W}c_w(s,\ell)a_0^w(b_0-a_0)^{W-w},
\end{gathered}
\label{eq:v9-22}
\end{equation}

where \(c_w(s,\ell)\) counts weight-\(w\) errors in the class. Indeed,

\begin{equation}
\begin{gathered}
M_W(s,\ell)=
\frac{(b_0-a_0)^{n-W}}{b_0^n}\,
\widetilde M_W(s,\ell),
\end{gathered}
\label{eq:v9-23}
\end{equation}

The common positive factor does not change which class maximizes the probability mass. The code-capacity evaluator compares integer class masses with a deterministic tie break, then converts the selected failure masses to floating-point probabilities. Full enumeration removes truncation error but still returns numerical probabilities. The measurement-noise evaluator instead accumulates and maximizes class masses in floating-point arithmetic (Sec.~\ref{a.-model-measured-checks-and-evaluation-bounds}). Neither implementation supplies a separately verified roundoff enclosure for its stored endpoints or floating-point operations.

The number of retained patterns per sector is
\(N_W=\sum_{w=0}^{W}\binom nw\), rather than \(2^n\).  We use \(W=4\) during
search. Candidates retained for final evaluation are scored by full
enumeration for \(n\le20\) and by the \(W=6\) bracket for \(n=30,40\).

\begin{figure*}[t]
\centering
\includegraphics[width=\linewidth]{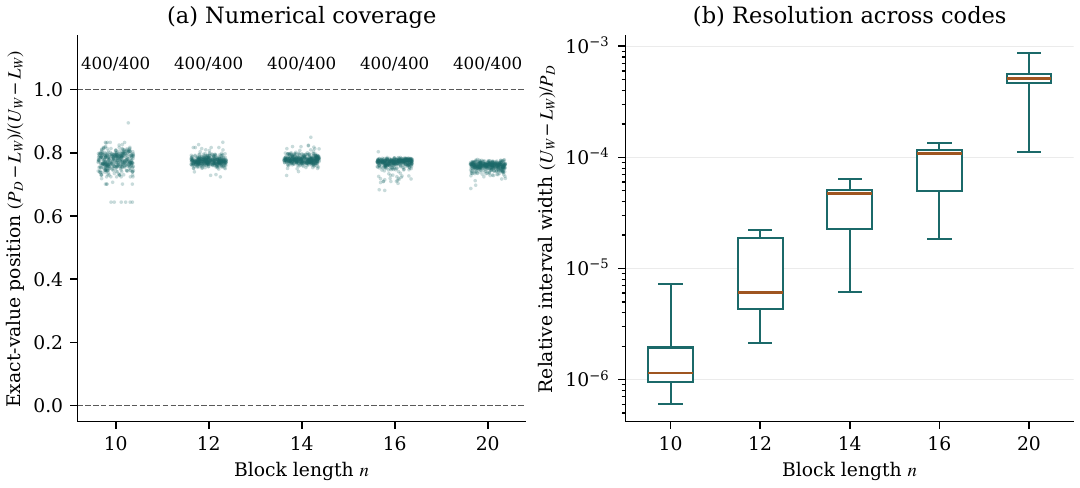}
\caption{Numerical validation and resolution of truncated reliability bounds. Independent \(X/Z\) noise has \(p=0.01\) per sector, with perfect syndrome extraction and sector-wise DML decoding; \(W=4\) in both panels. The validation set contains 400 codes at each of \(n=10,12,14,16,20\), including feasible and infeasible designs, and is distinct from the 477 retained candidates in the main experiment. (a) Each point locates an archived full-enumeration failure value within its truncation interval: 0 and 1 denote the lower and upper endpoints. Horizontal jitter only separates points. Counts report covered records; no violation is found at absolute tolerance \(10^{-14}\). (b) Relative interval widths use \(P_D\) as denominator. Boxes span the 25th to 75th percentiles, center lines mark medians, and whiskers show the full range across codes; these are not confidence intervals. All exact values and interval widths are positive. Proposition 1 establishes the general truncation bound; panel (a) checks its numerical implementation on the validation set.}
\label{fig:validation}
\end{figure*}

On an independent validation set of 2000 codes (400 at each of \(n=10,12,14,16,20\)), every full-enumeration value lies in its \(W=4\) interval at absolute tolerance \(10^{-14}\) [Fig.~\ref{fig:validation}(a)]. For final menus, the largest relative interval widths \((U_W-L_W)/L_W\) are approximately \(2.93\times10^{-3}\) in the main menus and \(3.03\times10^{-3}\) in the two \(n=40\) union menus. Appendix~\ref{app:validation-details} gives the validation and width definitions; threshold decisions use individual upper endpoints.

\subsection{Certified threshold
decisions}\label{d.-certified-threshold-decisions}

Let \(\mathcal A\) be a finite set of evaluated feasible codes, with
valid bounds \(L(C)\le P_D(C)\le U(C)\) and exact check costs
\(B_*(C)\). For a maximum allowed failure probability \(\tau\),
attainment means failure at or below \(\tau\). Define

\begin{equation}
\begin{gathered}
\mathcal A_\tau^{\mathrm{cert}}
=\{C\in\mathcal A:U(C)\le\tau\}.
\end{gathered}
\label{eq:v9-24}
\end{equation}

If this set is nonempty, the rule

\begin{equation}
\begin{gathered}
\widetilde C_\tau\in
\underset{C\in\mathcal A_\tau^{\mathrm{cert}}}
{\operatorname{arg\,min}}\ B_*(C),
\end{gathered}
\label{eq:v9-25}
\end{equation}

with ties resolved by smaller \(U(C)\), selects a code satisfying
\(P_D(\widetilde C_\tau)\le\tau\). It also minimizes check weight among
the candidates whose bounds certify this requirement. A code with
\(L(C)>\tau\) is certified to fail the requirement; when
\(L(C)\le\tau<U(C)\), its status remains unresolved by the interval.

 A cheaper code with \(P_D\le\tau<U\) remains possible, and ordering upper bounds need not order true failure probabilities. The selected code is Pareto-optimal for the upper-bound objectives \((U,B_*)\) within \(\mathcal A\), while true Pareto membership and minimum cost among all truly feasible candidates remain unresolved. Exact-value conclusions are recovered when the intervals collapse to true probabilities. An empty certified set means the current bounds provide no certified choice.

Similarly, if at least one candidate satisfies a budget \(b\), we can
use the conservative rule

\begin{equation}
\begin{gathered}
\widetilde C_b\in
\underset{C\in\mathcal A:\,B_*(C)\le b}
{\operatorname{arg\,min}}\ U(C),
\end{gathered}
\label{eq:v9-26}
\end{equation}

with ties resolved by smaller \(B_*\). This minimizes the available upper bound within budget and certifies \(P_D(\widetilde C_b)\le U(\widetilde C_b)\); the true-failure ranking remains unresolved.

\subsection{Dominance certified by
intervals}\label{e.-dominance-certified-by-intervals}

A sufficient condition for \(C'\) to dominate \(C\) in the true
objectives is

\begin{equation}
\begin{gathered}
B_*(C')\le B_*(C),\qquad U(C')\le L(C),
\end{gathered}
\label{eq:v9-27}
\end{equation}

with at least one of these inequalities strict. The probability ordering
follows from

\begin{equation}
\begin{gathered}
P_D(C')\le U(C')\le L(C)\le P_D(C).
\end{gathered}
\label{eq:v9-28}
\end{equation}

If the cost inequality is strict, equal interval endpoints suffice; at
equal costs, strict separation \(U(C')<L(C)\) is required by this test.
This rule certifies dominance and removes those candidates; overlapping intervals can leave true dominance among survivors unresolved.

We apply the test with search-time \(W=4\) brackets, refine survivors, and repeat the test; interval nesting preserves dominance established before refinement. The resulting set contains candidates not excluded by certified dominance within the searched pool. Surviving the test does not establish true Pareto optimality or search completeness.
\paragraph*{Scope of certification.} The two-sided bounds certify omitted probability mass under the stated noise and DML model; they are not statistical confidence intervals. The stored-endpoint roundoff limitation in Sec.~\ref{c.-evaluation-and-numerical-checks} applies throughout. Threshold and dominance conclusions concern specified finite evaluated sets, not the full feasible space. A menu minimum is established when the selected upper bound meets the requirement and every cheaper menu candidate has a lower bound above it; interval survival alone proves neither true Pareto membership nor search completeness. Search-time hypervolume instead uses truncated reliability coordinates to compare search outcomes.

\section{Search and evaluation
protocol}\label{iv.-search-and-evaluation-protocol}

We use a controlled search comparison to test how the reliability signal
affects the candidate sets found for the design problem in Sec.~\ref{ii.-decoder-aware-multiobjective-formulation}.
Three primary search arms share a decomposition simulated-annealing kernel and differ in the reliability term supplied to that kernel. Two additional
single-objective arms provide workflow-level comparisons. All five arms
are evaluated using the same decoder-based application metric,
irrespective of the signal driving their search. Execution details and checkpoint extraction are collected in Appendix~\ref{app:search-execution}.

\subsection{Common kernel and reliability
signals}\label{a.-common-kernel-and-reliability-signals}

The three primary arms use nine annealing chains with fixed weights
\(a_j=j/8\), \(j=0,\ldots,8\), and energy

\begin{equation}
\begin{gathered}
E_j(C)=a_jR(C)+(1-a_j)\widetilde B(C)
\\ +\lambda\,\operatorname{viol}(C),
\\
\widetilde B(C)=\frac{B_*(C)}{n(r_X+r_Z)},
\end{gathered}
\label{eq:v9-29}
\end{equation}

\begin{equation}
\begin{gathered}
\operatorname{viol}(C)=\max\{0,3-d(C)\}
+\#\{c:k_c=0\}.
\end{gathered}
\label{eq:v9-30}
\end{equation}

The main comparison sets \(\lambda=1\). Each proposal applies a CNOT
transformation between a uniformly sampled ordered pair of distinct
qubits. This preserves CSS commutation and the stabilizer ranks, while
distance and the component constraint are handled through the penalty
and the feasible archive. 

The reliability signals are

\begin{equation}
\begin{gathered}
R_{\mathrm{dec}}(C)=
\frac{\log_{10}[1+\hat P_4(C)/p^2]}
{\log_{10}\{1+[1-(1-p)^{2n}]/p^2\}},
\end{gathered}
\label{eq:v9-31}
\end{equation}
The normalization maps \(\hat P_4=0\) to zero and \(\hat P_4=1-(1-p)^{2n}\) to one; the logarithm compresses the range between these reference values.

\begin{equation}
\begin{gathered}
R_{\mathrm{dist}}(C)=1-\frac{d(C)-1}{n-1},
\\
R_{\mathrm{Perlin}}(C)=
\frac{\log m_{\mathrm{log}}(C)/\log1000-d(C)+n-1}{n}.
\end{gathered}
\label{eq:v9-32}
\end{equation}

Here \(m_{\mathrm{log}}(C)\) is the total number of weight-\(d(C)\)
nontrivial pure-type logical operators in the two sectors, counted by
the evaluator's logical weight spectra. In the results tables, the label \((d,m)\) abbreviates \((d,m_{\mathrm{log}})\); it does not
refer to the number of measured checks. The last signal is an affine
rescaling of the distance--multiplicity energy used in the
Perlin-inspired search \cite{perlin}. The distance-based signals use different normalizations: at fixed \(m_{\mathrm{log}}\), a unit increase in \(d\) changes \(R_{\mathrm{dist}}\) by \(-1/(n-1)\) and \(R_{\mathrm{Perlin}}\) by \(-1/n\).

For each seed, all arms receive the same initial population of 24 random CSS codes at the prescribed ranks. The primary arms use the same rule to initialize their nine chains; Appendix~\ref{app:search-execution} gives the chain schedule, acceptance rule, and temperature adaptation.

The primary arms share the initialization rule, chain weights, proposal operator, adaptation rule, constraint penalty, evaluator, and query allocation. Realized temperatures and trajectories can differ because
they depend on the signal-specific energy changes. Sharing a seed pairs runs with a common starting population; signal-specific energies can produce different trajectories.

The two additional arms minimize \(-d\) or
\(\log m_{\mathrm{log}}/\log1000-d\) on all nine chains. They share the
evaluator, initial population, proposal operator, and
temperature-adaptation procedure, but rank the initial population by
their own energy to select chain starts. Their acceptance energies
contain neither a check-cost term nor the component/distance penalty
used by the primary arms. Feasibility is still enforced when
constructing the application archive. These local single-objective baselines compare workflow choices, including differences in initialization and penalty treatment; they are not reproductions of published optimization workflows.

\subsection{Query accounting and candidate
archives}\label{b.-query-accounting-and-candidate-archives}

Each evaluator call counts as one query, including initialization, rejected proposals, and repeated codes. All arms use the same evaluator and obtain the same metrics for a given code, even when their search energies use only a subset of those metrics.

Equal query budgets control evaluator calls, including initialization; they are not wall-clock budgets. Final refinement of
pooled candidates, described in Sec.~\ref{iii.-certified-evaluation-and-decision-rules}, is performed separately from
the search budget.

Every evaluated feasible candidate is offered to the application
archive, including proposals rejected by the annealing chain. The
archive retains nondominated points in
\((R_{\mathrm{dec}},\widetilde B)\), with one representative per
objective point and a numerical comparison tolerance of \(10^{-12}\).
This search-time archive uses \(\hat P_4\).
Across arms, search-time archives collect nondominated evaluated candidates. Pooling these archives, deduplicating codes, and applying interval-dominance filtering yields the menus used for selection.

\subsection{Cells, seeds, and the frozen
design}\label{c.-cells-seeds-and-the-frozen-design}

The main experiment comprises eight cells, \(n\in\{14,20,30,40\}\) and
\(k\in\{2,4\}\), with \(r_X=r_Z=(n-k)/2\) and \(p=0.01\). Each of the
five arms is run with the same 30 seeds,

\begin{equation}
\begin{gathered}
\mathrm{seed}_i=1009+37i,\qquad i=0,\ldots,29.
\end{gathered}
\label{eq:v9-33}
\end{equation}

The resulting design contains \(8\times5\times30=1200\) runs. The
per-run query budgets are given in Table~\ref{tab:budget}.

\begin{table}[t]
\caption{Per-run query budgets for the main experiment.}
\label{tab:budget}
\centering
\begin{tabular}{@{}p{.29\columnwidth}p{.29\columnwidth}p{.29\columnwidth}@{}}
\toprule

Physical qubits \(n\) & Logical qubits \(k\) & Query budget per run \\
\midrule

14 & 2, 4 & 5,000 \\
20 & 2, 4 & 20,000 \\
30 & 2, 4 & 20,000 \\
40 & 2, 4 & 40,000 \\
\bottomrule
\end{tabular}
\end{table}

Settings were fixed after a 108-run tuning phase. Two tuning seeds (1009 and 1046) also occur in the main runs, so this comparison is not wholly held out (Appendix~\ref{app:search-execution}). 

The main experiment includes all 1200 planned cell--arm--seed runs, with
matching seed sets across arms and final checkpoints at the prescribed
budgets. All completed valid runs are included. Runs that find no feasible code contribute zero application
hypervolume.

\subsection{A common application
hypervolume}\label{d.-a-common-application-hypervolume}

For a run at query count \(t\), let \(\mathcal A_t\) denote its feasible
application archive. Its hypervolume \cite{zitzler} is

\begin{equation}
\begin{gathered}
\operatorname{HV}(\mathcal A_t)=
\operatorname{area}\!\left[
\bigcup_{C\in\mathcal A_t}
[R_{\mathrm{dec}}(C),1.01]\times
[\widetilde B(C),1.01]
\right],
\end{gathered}
\label{eq:v9-34}
\end{equation}

where points outside the reference rectangle contribute nothing and the
empty archive has hypervolume zero. Larger HV indicates a larger area
dominated by the candidate set in these two minimization coordinates.
The reference point is fixed at \((1.01,1.01)\) for every arm.

All arms are scored in \((R_{\mathrm{dec}},\widetilde B)\),
including arms driven by distance or multiplicity. Evaluating each arm
in its own search coordinates would compare different objectives. This
common metric tests candidate quality for the decoder-based design task.
 Its
numerical value depends on the normalization and reference point, so
performance comparisons are made within each \((n,k)\) cell at fixed
\(p\).

The primary outcome is final per-run HV. Feasibility and mean HV at 25\%, 50\%, 75\%, and 100\% of the query budget describe search progress; the means include all 30 seeds and retain zero-valued failures. Appendix~\ref{app:checkpoints} specifies how these checkpoints are extracted.

\subsection{Paired inference and multiplicity
correction}\label{e.-paired-inference-and-multiplicity-correction}

For each cell and comparator \(a\), we form 30 paired differences

\begin{equation}
\begin{gathered}
\Delta_i=\operatorname{HV}_{\mathrm{dec},i}
-\operatorname{HV}_{a,i}.
\end{gathered}
\label{eq:v9-35}
\end{equation}

The unit of replication is the seeded search run, not an archived code
or a trajectory checkpoint. We report the mean difference and the number
of positive and negative pairs, using a \(10^{-12}\) tolerance for the
latter. Statistical comparisons use a two-sided sign-flip test of
\(|\overline\Delta|\), with 100,000 independent random sign vectors. With \(T_r=|30^{-1}\sum_i\epsilon_{ri}\Delta_i|\)
and equiprobable signs \(\epsilon_{ri}\in\{-1,1\}\), the Monte Carlo p
value is

\begin{equation}
\begin{gathered}
p_{\mathrm{raw}}=
\frac{1+\#\{r:T_r\ge|\overline\Delta|-10^{-15}\}}
{100001}.
\end{gathered}
\label{eq:v9-36}
\end{equation}

The sign-flip test assumes sign exchangeability of paired differences under the null and concerns the seeded search comparison, not a sample of all CSS codes.

We apply a joint Holm correction across all 32 tests \cite{holm}. This is at least as stringent as the separate corrections specified for two 16-test families; the main-comparison significance classifications at the 0.05 level are unchanged under that original plan (Appendix~\ref{app:search-execution}). A nonsignificant result means that the test did not detect a difference; it establishes neither equivalence nor noninferiority.

\subsection{Secondary analyses}\label{f.-secondary-analyses-and-data-provenance}

Five sensitivity campaigns vary the query budget at \(n=40\), the penalty weight, or the physical error probability; their results are reported separately in Appendix~\ref{app:sensitivity}. Candidate pooling for decision
examples is also separate from the per-run HV comparison: pooled menus
describe available choices, while per-run HV measures search
performance.

Exploratory search-time threshold-cost screening uses \(\hat P_4\le\tau\); thresholds for \((14,4)\) and \((20,4)\) were chosen post hoc. Certified selection in Sec.~\ref{b.-selecting-codes-with-a-failure-requirement} requires \(U(C)\le\tau\).

\section{Results}\label{v.-results}

We first examine the reliability--check-weight choices produced by the
search and apply the upper-bound decision rule of Sec.~\ref{iii.-certified-evaluation-and-decision-rules}. We then
compare the search signals using the paired, per-run hypervolume metric
of Sec.~\ref{iv.-search-and-evaluation-protocol}. These analyses address different questions: pooling
candidates identifies available design choices, whereas paired runs
measure how effectively each signal explores the design space.

\begin{table}[htbp]
\caption{Candidate menus from the main experiment at
\(p=0.01\). Counts refer to candidates retained by interval-dominance
filtering. The two \(n=40\) union menus, discussed below, are not
included in this table.}
\label{tab:r1}
\centering
\begin{tabular}{@{}lrr@{}}
\toprule

\((n,k)\) & Retained candidates & Range of \(B_*\) \\
\midrule

(14,2) & 13 & 38--56 \\
(14,4) & 4 & 46--50 \\
(20,2) & 36 & 56--100 \\
(20,4) & 40 & 56--101 \\
(30,2) & 61 & 84--190 \\
(30,4) & 76 & 87--189 \\
(40,2) & 117 & 116--293 \\
(40,4) & 130 & 114--302 \\
\bottomrule
\end{tabular}
\end{table}

\subsection{Reliability--check-weight
trade-offs}\label{a.-reliabilitycheck-weight-trade-offs}

Together, the eight main menus contain 477 retained candidates; we call this aggregate the main pool used for the measurement-noise analysis. These candidates have full-enumeration failure estimates for \(n\le20\) and weight-6 intervals for \(n=30,40\), after
the refinement and interval-dominance filtering described in Sec.~\ref{iii.-certified-evaluation-and-decision-rules}.
Table~\ref{tab:r1} summarizes their counts and check-weight ranges.

\begin{figure*}[t]
\centering
\includegraphics[width=\linewidth]{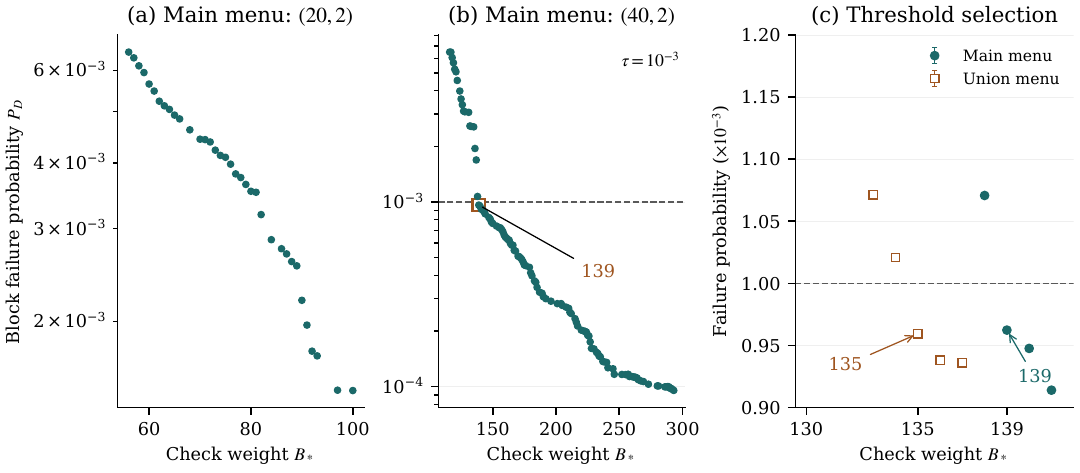}
\caption{Reliability--check-weight menus and threshold selection at \(p=0.01\) per sector (independent \(X/Z\) noise, perfect syndrome extraction, sector-wise DML). (a) All 36 entries of the \((20,2)\) main menu, by full enumeration. (b) All 117 entries of the \((40,2)\) main menu, with \(W=6\) bounds; dashed line, \(\tau=10^{-3}\); outlined point, selected weight 139. (c) The \((40,2)\) threshold region: filled circles, main menu; open squares, union menu after an additional 80,000-query campaign, selecting weight 135. Symbols mark interval midpoints and bars span \([L,U]\); panels (a,b) use log failure axes and (c) a restricted linear axis. Check weight is the minimum total data-qubit participation over independent pure-type check bases.}
\label{fig:menus}
\end{figure*}

Figure~\ref{fig:menus}(a,b) displays the main-menu trade-offs for \((20,2)\) and \((40,2)\). At \((n,k)=(40,2)\), the lowest-cost main-menu candidate has \(B_*=116\) and failure interval approximately \([6.48614,6.48642]\times10^{-3}\). The candidate with the smallest failure lower bound has \(B_*=293\) and interval \([9.51738,9.54530]\times10^{-5}\). Thus the menu offers substantially different reliability levels at different check weights while keeping the encoding resources fixed. Symbols mark interval midpoints; vertical bars span the stored bounds and can be shorter than the markers. 

The extent of the observed trade-off varies across cells. At \((14,4)\), only four candidates remain, with \(B_*=46\)--50 and full-enumeration failure values spanning approximately \(1.2212\times10^{-2}\) to \(1.2645\times10^{-2}\). This narrow menu contrasts with the broader range at \((40,2)\). Minimum observed costs also vary across cells: the main pool includes \(B_*=38\) at \((14,2)\) and \(B_*=56\) in both \(n=20\) cells. 

\subsection{Selecting codes with a failure
requirement}\label{b.-selecting-codes-with-a-failure-requirement}

To illustrate code selection, we apply the rule \(U(C)\le\tau\)
to the stored \((40,2)\) menus. Table~\ref{tab:r2} reports the minimum check
weight among candidates passing this test, with smaller \(U\) breaking
cost ties. The thresholds \(10^{-3}\), \(5\times10^{-4}\), and \(10^{-4}\) provide descriptive examples of increasingly stringent requirements.

\begin{table*}[t]
\caption{Upper-bound threshold selections for \((n,k)=(40,2)\)
at \(p=0.01\). The main menu uses the 40,000-query experiment; the union
menu combines retained candidates from that experiment and the separate
80,000-query campaign. All decisions use unrounded stored endpoints; displayed intervals are rounded for readability.}
\label{tab:r2}
\centering
\begin{tabular}{@{}
  p{(\linewidth - 6\tabcolsep) * \real{0.2308}}
  p{(\linewidth - 6\tabcolsep) * \real{0.2308}}
  p{(\linewidth - 6\tabcolsep) * \real{0.3077}}
  p{(\linewidth - 6\tabcolsep) * \real{0.2308}}@{}}
\toprule

\begin{minipage}[b]{\linewidth}\raggedright
Failure requirement \(\tau\)
\end{minipage} & \begin{minipage}[b]{\linewidth}\raggedright
Menu
\end{minipage} & \begin{minipage}[b]{\linewidth}\raggedleft
Selected \(B_*\)
\end{minipage} & \begin{minipage}[b]{\linewidth}\raggedright
Stored \([L,U]\), displayed approximately
\end{minipage} \\
\midrule

\(10^{-3}\) & Main & 139 & \([9.621890,9.624680]\times10^{-4}\) \\
\(10^{-3}\) & Union & 135 & \([9.590809,9.593600]\times10^{-4}\) \\
\(5\times10^{-4}\) & Main & 172 &
\([4.971169,4.973960]\times10^{-4}\) \\
\(5\times10^{-4}\) & Union & 170 &
\([4.944313,4.947104]\times10^{-4}\) \\
\(10^{-4}\) & Main & 286 & \([9.948635,9.976552]\times10^{-5}\) \\
\(10^{-4}\) & Union & 281 & \([9.860755,9.888672]\times10^{-5}\) \\
\bottomrule
\end{tabular}
\end{table*}

Figure~\ref{fig:menus}(c) resolves the \((40,2)\) threshold region. The main-menu code selected at \(\tau=10^{-3}\) has \(B_*=139\) and \(U\simeq9.62468\times10^{-4}\). Every cheaper entry in the full main menu has a lower bound above the threshold; the smallest such lower bound is approximately \(1.07067\times10^{-3}\). The interval data therefore establish both threshold attainment for the selected candidate and exclusion of cheaper entries in this menu, within the certification scope of Sec.~\ref{iii.-certified-evaluation-and-decision-rules}. Open squares separately show the union menu, whose threshold selection reduces the required check weight from 139 to 135. This improvement reflects an enlarged candidate pool, not a paired estimate of the performance gain of one search arm.

In all six examples, the selected upper bound is below the requirement and every cheaper entry in the corresponding menu has a lower bound above it. These interval comparisons establish minimum cost among truly threshold-feasible entries of that menu. 

\subsection{Effect of the reliability
signal}\label{c.-effect-of-the-reliability-signal}

\begin{figure*}[t]
\centering
\includegraphics[width=\linewidth]{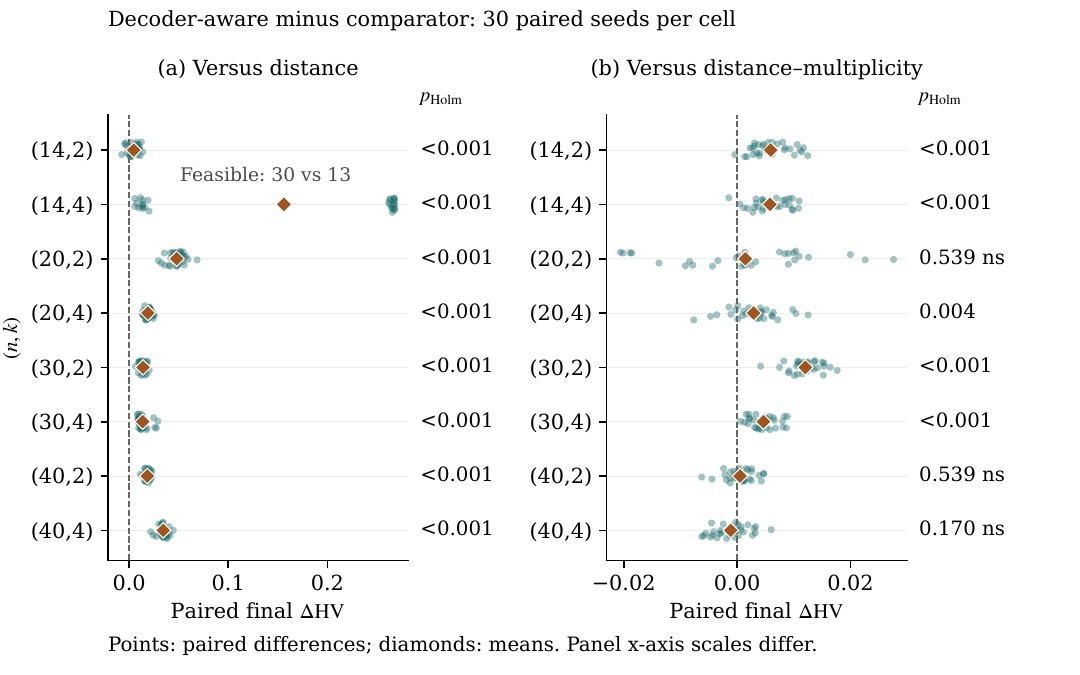}
\caption{Paired effects of the reliability signal in a common search kernel (\(p=0.01\) per independent \(X/Z\) sector). Points show final application-HV differences between decoder-aware search and (a) distance or (b) distance--multiplicity for the same seed and cell; diamonds mark means of 30 pairs, and jitter separates points. Query budgets are in Table~\ref{tab:budget}. HV uses common \(W=4\) truncated coordinates and the reference point of Sec.~\ref{iv.-search-and-evaluation-protocol}; it measures search performance. Axes differ between panels. Displayed p values use the joint 32-test Holm correction; ns denotes \(p_{\rm Holm}\ge0.05\). Runs without a feasible candidate score zero.}
\label{fig:paired}
\end{figure*}

Figure~\ref{fig:paired}(a) shows that decoder-aware guidance produces significantly higher mean final application HV than distance guidance in all eight main cells (Holm-adjusted \(p<10^{-3}\)). Seven cells have mean paired gains between 0.00463 and 0.04793; the \((14,4)\) gain is 0.15607. In that cell, decoder-aware search reaches feasibility in 30 of 30 runs, versus 13 of 30 for distance. Because runs without a feasible code score zero, this contrast reflects access to feasible designs as well as the quality of menus found.

\begin{table*}[t]
\caption{Paired final-HV differences, decoder-aware minus the
\((d,m)\) signal, for the main experiment. Each row contains 30 seed
pairs. Adjusted p values use the actual joint Holm correction across 32
main-experiment comparisons, as documented in Sec.~\ref{iv.-search-and-evaluation-protocol}. Displayed values
are rounded. The repeated 0.000320 entries are the minimum Holm-adjusted value at the 100,000-vector Monte Carlo resolution: \(32/100001\), rounded.}
\label{tab:r3}
\centering
\begin{tabular}{@{}
  p{(\linewidth - 6\tabcolsep) * \real{0.2000}}
  p{(\linewidth - 6\tabcolsep) * \real{0.2667}}
  p{(\linewidth - 6\tabcolsep) * \real{0.2667}}
  p{(\linewidth - 6\tabcolsep) * \real{0.2667}}@{}}
\toprule

\begin{minipage}[b]{\linewidth}\raggedright
\((n,k)\)
\end{minipage} & \begin{minipage}[b]{\linewidth}\raggedleft
Mean \(\Delta\mathrm{HV}\)
\end{minipage} & \begin{minipage}[b]{\linewidth}\raggedleft
Positive/negative pairs
\end{minipage} & \begin{minipage}[b]{\linewidth}\raggedleft
Adjusted p
\end{minipage} \\
\midrule

(14,2) & +0.005929 & 29/1 & 0.000320 \\
(14,4) & +0.005807 & 29/1 & 0.000320 \\
(20,2) & +0.001465 & 19/11 & 0.538635 \\
(20,4) & +0.002931 & 24/6 & 0.003560 \\
(30,2) & +0.012042 & 30/0 & 0.000320 \\
(30,4) & +0.004648 & 30/0 & 0.000320 \\
(40,2) & +0.000518 & 18/12 & 0.538635 \\
(40,4) & \ensuremath{-}0.001114 & 11/19 & 0.169828 \\
\bottomrule
\end{tabular}
\end{table*}

Five cells show a positive difference detected at the 0.05 level; the
remaining three do not. At \((40,4)\), the decoder-aware and \((d,m)\)
mean HV values are 0.8008 and 0.8019, respectively. The negative difference is nonsignificant, and the advantage is not uniform across cells. The three nonsignificant comparisons provide no evidence of equivalence.

The pooled candidates also show that the two signals can contribute
different options. In the main \((40,2)\) menu, 71 retained candidates
are attributed to the decoder-aware arm and 46 to the \((d,m)\) arm. At
\((40,4)\) the corresponding counts are 51 and 69, with the remaining
ten attributed to other arms. These counts identify which search arm contributed each retained candidate.

The two local single-objective baselines have lower mean HV than the
decoder-aware arm in all 16 main comparisons, with adjusted
\(p<10^{-3}\). Their mean paired deficits range from approximately
0.00606 to 0.08781. These 16 comparisons evaluate the local single-objective workflows defined in Sec.~\ref{iv.-search-and-evaluation-protocol}.

\subsection{Search-budget dependence and the union
menus}\label{d.-search-budget-dependence-and-the-union-menus}

The trajectory data do not support a fixed ordering of the three signals throughout the search. At \((40,2)\), the \((d,m)\) mean HV exceeds the decoder-aware mean at the 25\%, 50\%, and 75\% checkpoints, before the small positive decoder-aware difference at the final budget. In the separate 80,000-query campaign, the decoder-aware minus \((d,m)\) mean difference is positive and detected by the raw sign-flip test at \((40,2)\), but remains negative and nonsignificant at \((40,4)\). These comparisons are within each campaign; changes in the between-arm effect across budgets were not tested. Appendix~\ref{app:budget-results} gives the full numerical results.

Longer searches affect both the comparison between signals and the menus available for selection. Paired per-run HV measures search performance; pooling and filtering measure the menu available for selection. Merging the main and longer-campaign retained sets, deduplicating their code representations, and applying the interval-dominance test gives union menus of 111 and 137 candidates for \((40,2)\) and \((40,4)\), respectively, compared with 117 and 130 in the main menus. New candidates can dominate several older entries, so an improved menu can contain fewer retained candidates. The main eight-cell menu contains 477 candidates; replacing its two \(n=40\) menus with the union menus gives 478. Union menus are used only where explicitly identified. The measurement-noise analysis in the following section uses the original 477 candidates.

\subsection{Penalty and physical-error
sensitivity}\label{e.-penalty-and-physical-error-sensitivity}

The penalty and physical-error analyses cover only \((20,2)\) and \((30,2)\), with three signals and 30 paired seeds per cell. Their p values are unadjusted and are separate from the main 32-test Holm family. Changing \(\lambda\) from 1 to 0.3 or 3 leaves final HV for every paired seed unchanged for the decoder-aware and \((d,m)\) arms in these cells. This limited insensitivity is specific to the two tested cells and does not imply equivalence to a hard constraint. Appendix~\ref{app:parameter-results} reports the complete results, including the distance arm.

At \(p=0.001\), the decoder-aware signal has positive mean differences from \((d,m)\) in both tested cells. At \(p=0.03\), the difference is slightly negative and nonsignificant at \((20,2)\), but remains positive at \((30,2)\). Its mean HV is higher than that of distance in all ten cell--variant comparisons, with raw \(p<10^{-3}\). Comparisons remain within each physical-error setting because both the objective and its HV normalization depend on \(p\).

These comparisons show how search performance depends on the signal, budget, and data-error setting. The next section changes a different assumption---perfect syndrome measurements---and examines the resulting choices within the fixed main pool.

\section{Extension to noisy syndrome
measurements}\label{vi.-extension-to-noisy-syndrome-measurements}

The reliability--weight menus of Sec.~\ref{v.-results} assume perfect syndrome
measurements. We now re-evaluate the 477 codes in the main menus to ask
how independent measurement errors change their ordering and the
allocation of a check-weight budget. This extension keeps the code pool fixed and adds a choice of redundant checks, so it tests design decisions within the available menus. Repeated syndrome extraction provides a standard route to handling measurement faults in topological memories \cite{dennis2002}. Here we instead study redundant checks within a strict single-round model, following the data-syndrome perspective \cite{ashikhmin}.

\subsection{Model, measured checks, and evaluation
bounds}\label{a.-model-measured-checks-and-evaluation-bounds}

For each error sector \(\alpha\in\{X,Z\}\), the decoder observes \begin{equation}
\begin{gathered}
s'=Me+f\pmod 2,
\end{gathered}
\label{eq:v9-37}
\end{equation} where \(e\) and \(f\) are column vectors, consistent with Sec.~\ref{ii.-decoder-aware-multiobjective-formulation}.
The measured matrix \(M\) contains checks of the opposite Pauli type to
the error sector: \(Z\)-type checks for \(X\) errors and \(X\)-type
checks for \(Z\) errors. The data-error bits in \(e\) are independent
Bernoulli variables with probability \(p=0.01\), and the
measurement-error bits in \(f\) are independent Bernoulli variables with
probability \(q\in\{0.01,0.005\}\). Data and measurement errors, as well
as the two sectors, are independent. The decoder infers the class of the
data error modulo the corresponding stabilizer space, including its true
basis syndrome and logical class. Recovery fails whenever the residual data error is not a stabilizer; residual syndrome errors therefore count as failures. 

For each check space, the measured rows consist of a minimum-weight
basis followed by the \(t\) lightest unused nonzero elements of that
space. Ties are broken by integer row label (Appendix~\ref{app:noise-checks}). The same \(t\) is used in both sectors, whose ranks in
this data set are \(r=(n-k)/2\). We evaluate the distinct values in \begin{equation}
\begin{gathered}
\mathcal T=\{0,1,2,4,8,r\},\\
 B_{\mathrm{meas}}(C,t)=\sum_{a\in M_X(C,t)}\mathrm{wt}(a)
           \\ +\sum_{a\in M_Z(C,t)}\mathrm{wt}(a).
\end{gathered}
\label{eq:v9-38}
\end{equation} Here \(C\) denotes a code, and \(B_{\mathrm{meas}}(C,0)=B_*(C)\). The
maximum tested redundancy is \(t_{\max}=\max(8,r)\): it equals 8 for the
two 14-qubit cells and \(r\) for the larger cells. Even at this maximum,
the measured set contains only \(r+t_{\max}\) rows per sector, not every
nonzero stabilizer. The measured weight is a resource proxy;
measurement-error probability is independent of row weight in this
model.

The evaluation enumerates fault patterns with total weight
\(|e|+|f|\le6\). For error sector \(\alpha\) with \(m_\alpha\) measured
rows (\(m_X=m_Z=r+t\) here), define the omitted probability by \begin{equation}
\begin{gathered}
T_{\alpha,W}:=T_W(n,m_\alpha;p,q)\\ =1-\!\sum_{\substack{0\le a\le n,\;0\le b\le m_\alpha\\a+b\le W}}
 {n\choose a}p^a(1-p)^{n-a}
\\ {}\times {m_\alpha\choose b}q^b(1-q)^{m_\alpha-b},\\ W=6.
\end{gathered}
\label{eq:v9-39}
\end{equation} For \(p=q\), this reduces to the binomial tail on \(n+m_\alpha\)
bits; otherwise it is the tail of the sum of two binomial variables.
Applying the truncated class-mass argument of Sec.~\ref{iii.-certified-evaluation-and-decision-rules} gives sector
bounds
\(L_{\alpha,W}^{\mathrm{ds}}\le P_\alpha^{\mathrm{ds}}\le U_{\alpha,W}^{\mathrm{ds}}=\min(1,L_{\alpha,W}^{\mathrm{ds}}+T_{\alpha,W})\).
Here \(L_{\alpha,W}^{\mathrm{ds}}\) is the unnormalized truncated
failure mass in that sector. Independence then gives the block bounds \begin{equation}
\begin{gathered}
L_{\rm ds}=1-(1-L_{X,W}^{\mathrm{ds}})(1-L_{Z,W}^{\mathrm{ds}}),\\
 U_{\rm ds}=1-(1-U_{X,W}^{\mathrm{ds}})(1-U_{Z,W}^{\mathrm{ds}}).
\end{gathered}
\label{eq:v9-40}
\end{equation} We denote the stored truncated block value by \(\widehat P_{\rm ds}=L_{\rm ds}\). This is an unnormalized lower endpoint, not an unbiased statistical estimate or the full-distribution failure probability of an explicitly simulated decoder. The class-mass argument bounds full-distribution degenerate-ML failure. Stored endpoints retain the numerical qualification of Sec.~\ref{c.-evaluation-and-numerical-checks}.

The evaluation comprises 5,644 noisy-syndrome cases for the fixed main pool, both measurement-error settings, and the specified redundancy grids. Separate zero-measurement-noise checks recover the stored code-capacity lower endpoints to numerical precision; Appendix~\ref{app:noise-checks} reports these implementation checks, which are not additional independent code samples.

\subsection{Code ordering and the effect of
redundancy}\label{b.-code-ordering-and-the-effect-of-redundancy}

At \(q=p\), the cell medians of basis-only \(\widehat P_{\rm ds}\) range from 0.1068 to 0.1744. At maximum tested redundancy, cell medians of the within-code ratio of stored \(\widehat P_{\rm ds}\) values to their basis-only values range from 0.145 to 0.361.

Basis-only noisy failure remains strongly rank-correlated with the code-capacity value (Spearman \(\rho=0.906\)--1.000), yet the point-value nondominated sets change. In the six cells with \(n\ge20\), only 23.8\%--60.0\% of the code records remain nondominated under the point-value comparison; all four records remain at \((14,4)\).  Appendix~\ref{app:noise-ordering} gives the full summaries and rank correlations. Strong rank correlation therefore does not preserve the observed point-value trade-off sets. This motivates comparing code-first selection with joint code-and-redundancy selection under the same measured-weight budget. The results in Sec.~\ref{d.-what-the-intervals-establish} test whether selected designs show a strict gain.

\subsection{Allocating a measured-weight
budget}\label{c.-allocating-a-measured-weight-budget}

We compare two selection rules at each distinct measured-weight cost present in a cell's evaluated menu. The code-first rule chooses the code with the smallest stored code-capacity lower endpoint among those with \(B_*\le b\), then chooses the tested redundancy level that minimizes \(\widehat P_{\rm ds}\) within budget \(b\). The joint rule minimizes \(\widehat P_{\rm ds}\) over all evaluated \((C,t)\) with \(B_{\mathrm{meas}}(C,t)\le b\). Both rules therefore use the same finite menu, but optimize different decisions.

\begin{figure*}[t]
\centering
\includegraphics[width=\linewidth]{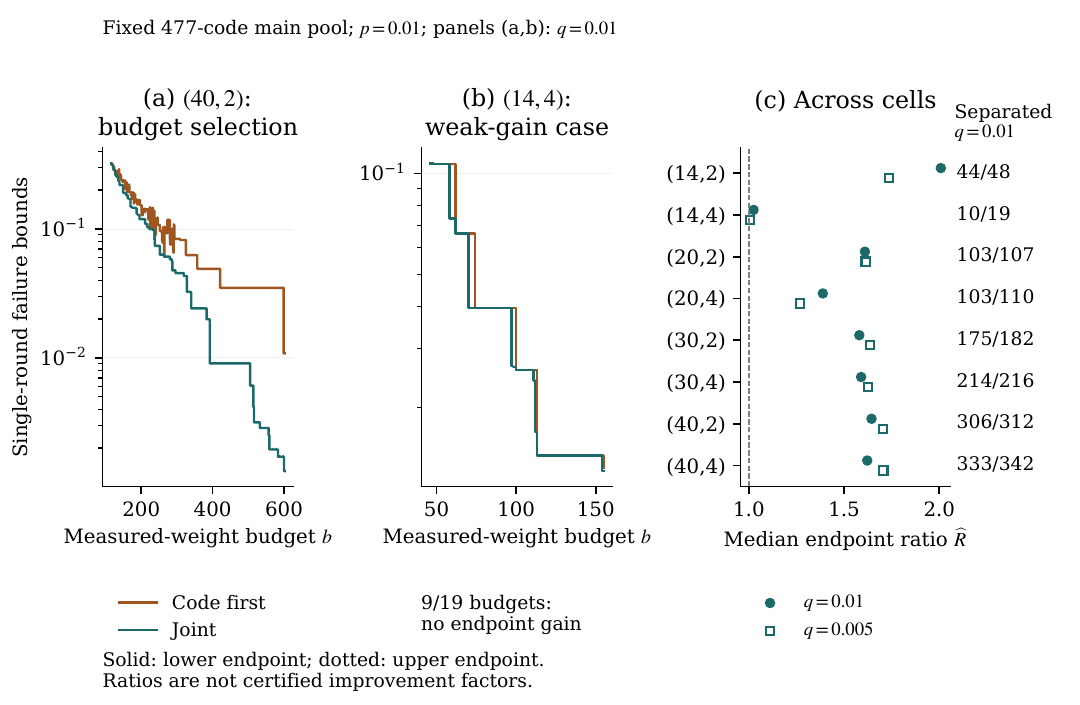}
\caption{Joint code--redundancy selection under a measured-weight budget for the fixed 477-code main pool (\(p=0.01\) per sector, strict single-round model, no ideal residual-error cleanup). The redundancy grid is \(\{0,1,2,4,8,r\}\), with \(t_{\max}=\max(8,r)\). (a,b) Code-first and joint selections at \(q=0.01\) for \((40,2)\) and \((14,4)\); solid steps are selected lower endpoints and dotted steps are \(W=6\) upper endpoints. At \((14,4)\), nine of 19 budgets show no endpoint gain. (c) Median endpoint ratios \(\widehat R=L_{\rm ds}(a_b)/L_{\rm ds}(j_b)\) for \(q=0.01\) and 0.005; right-hand counts give budgets with \(U_{\rm ds}(j_b)<L_{\rm ds}(a_b)\). Ratios are not certified true gain factors; interval separation supports the direction for selected designs under the stated numerical scope.}
\label{fig:noisebudget}
\end{figure*}

Ties are defined by exact equality of stored endpoints, without an additional tolerance. The code-first rule resolves equal code-capacity endpoints by the smaller archived code index within the cell. In its redundancy step and in the joint rule, ties between noisy endpoints are resolved by smaller measured-check weight, then smaller code index, then smaller redundancy level \(t\). Table~\ref{tab:m2} and Fig.~\ref{fig:noisebudget} use these tie-breaking rules. 

For the selected designs \(a_b\) and \(j_b\), respectively, define \begin{equation}
\begin{gathered}
\widehat R(b)=\frac{L_{\rm ds}(a_b)}{L_{\rm ds}(j_b)}.
\end{gathered}
\label{eq:v9-41}
\end{equation} The joint rule has \(\widehat R(b)\ge1\) by construction because its
feasible set contains the code-first choice. The empirical question is
the size and frequency of the gap. Table~\ref{tab:m2} reports medians with equal
weight on each distinct observed cost, rather than on every integer
budget or on independent experimental repetitions.

\begin{table*}[t]
\caption{Budget-allocation comparisons. The ratio columns
summarize truncated endpoint ratios. The final column separately counts
budgets at \(q=0.01\) for which \(U_{\rm ds}(j_b)<L_{\rm ds}(a_b)\),
certifying a strict comparison of the two selected designs subject to
the stored-endpoint numerical qualification in Sec.~\ref{a.-model-measured-checks-and-evaluation-bounds}.}
\label{tab:m2}
\centering
\begin{tabular}{@{}
  p{(\linewidth - 8\tabcolsep) * \real{0.1579}}
  p{(\linewidth - 8\tabcolsep) * \real{0.2105}}
  p{(\linewidth - 8\tabcolsep) * \real{0.2105}}
  p{(\linewidth - 8\tabcolsep) * \real{0.2105}}
  p{(\linewidth - 8\tabcolsep) * \real{0.2105}}@{}}
\toprule

\begin{minipage}[b]{\linewidth}\raggedright
\((n,k)\)
\end{minipage} & \begin{minipage}[b]{\linewidth}\raggedleft
Distinct budgets
\end{minipage} & \begin{minipage}[b]{\linewidth}\raggedleft
Median \(\widehat R\), \(q=0.01\)
\end{minipage} & \begin{minipage}[b]{\linewidth}\raggedleft
Median \(\widehat R\), \(q=0.005\)
\end{minipage} & \begin{minipage}[b]{\linewidth}\raggedleft
Strict interval-separated comparisons, \(q=0.01\)
\end{minipage} \\
\midrule

(14,2) & 48 & 2.0082 & 1.7371 & 44/48 \\
(14,4) & 19 & 1.0248 & 1.0063 & 10/19 \\
(20,2) & 107 & 1.6092 & 1.6132 & 103/107 \\
(20,4) & 110 & 1.3885 & 1.2682 & 103/110 \\
(30,2) & 182 & 1.5801 & 1.6363 & 175/182 \\
(30,4) & 216 & 1.5895 & 1.6261 & 214/216 \\
(40,2) & 312 & 1.6442 & 1.7047 & 306/312 \\
(40,4) & 342 & 1.6217 & 1.7062 & 333/342 \\
\bottomrule
\end{tabular}
\end{table*}

Figure~\ref{fig:noisebudget}(a) illustrates the budget-dependent choices at \((40,2)\), while panel (b) shows the weak-gain case \((14,4)\). Solid steps show the selected lower endpoints and dotted steps their upper endpoints; the steps represent choices from the evaluated discrete menu. Code-first selection can switch to a different code as the budget increases, so its noisy endpoint need not decrease monotonically.

Panel (c) separates the median endpoint ratios from the interval-separated comparison counts in Table~\ref{tab:m2}. At \(q=p\), median endpoint ratios lie between 1.3885 and 1.6442 for \(n\ge20\). The 14-qubit cells differ: the median is 2.0082 at \((14,2)\), but only 1.0248 at \((14,4)\), where the two rules return equal truncated values at nine of 19 budgets. At \(q=p/2\), the latter median falls to 1.0063. Benefits therefore depend on the menu and noise setting. The largest endpoint ratio, 49.34 at \(q=p\) in \((40,4)\), occurs at one budget on the tested redundancy grid; the cell medians above summarize the budget-wise pattern.

\subsection{What the intervals
establish}\label{d.-what-the-intervals-establish}

The reported endpoint ratios describe stored endpoints; the interval bounds below quantify the corresponding true-failure ratios. For each pair of selected
designs, positive endpoints instead give \begin{equation}
\begin{gathered}
\frac{L_{\rm ds}(a_b)}{U_{\rm ds}(j_b)}
 \le \frac{P_{\rm ds}(a_b)}{P_{\rm ds}(j_b)}
 \le \frac{U_{\rm ds}(a_b)}{L_{\rm ds}(j_b)}.
\end{gathered}
\label{eq:v9-42}
\end{equation} The comparisons in Table~\ref{tab:m2} and Fig.~\ref{fig:noisebudget}(c) use the stored bounds and the tie rules in Sec.~\ref{c.-allocating-a-measured-weight-budget}. At both \(q=p\) and \(q=p/2\), every budget with a strictly smaller joint-rule lower endpoint also satisfies \(U_{\rm ds}(j_b)<L_{\rm ds}(a_b)\). Each setting has two budgets with tied joint minima; using the largest upper endpoint across all tied designs leaves these counts unchanged. Subject to the numerical qualification in Sec.~\ref{a.-model-measured-checks-and-evaluation-bounds}, the intervals therefore establish the direction of these selected-design comparisons. Equation~\eqref{eq:v9-42} bounds the corresponding true gain; the optimality of either selected design over the full menu remains unresolved.

The largest relative interval width
\((U_{\rm ds}-L_{\rm ds})/L_{\rm ds}\) among the noisy evaluations is
approximately 3.32\% at \(q=p\), and 1.19\% at \(q=p/2\). These widths
are deterministic truncation bounds, not confidence intervals. They also
show why code-capacity interval precision cannot be assumed to carry
over unchanged to the enlarged fault space.

The extension identifies a concrete additional design variable: how a
fixed measured-weight budget is divided between the code's basis and
redundant checks. Its conclusions apply to this fixed pool, the
tested low-weight check sets, independent errors, and the strict
single-round DML criterion. 

\section{Discussion}\label{vii.-discussion}

\subsection{Design choices and what certification adds}\label{a.-decoder-aware-design-as-a-resource-constrained-choice}

We formulate CSS code design to connect a specified decoding task to a resource-constrained choice. At fixed encoding resources, the pair \((P_D,B_*)\) separates sector-wise degenerate maximum-likelihood failure from the minimum weight of an independent pure-type check basis. Keeping these objectives separate allows a candidate menu to serve different reliability requirements or check-weight budgets. Changing a requirement selects a different available design without redefining the objectives; the scalarizations used to search for candidates are a separate computational choice.

Basis minimization prevents generator changes alone from appearing as cost improvements. The component constraint excludes stabilizer-state factors while allowing direct sums that encode information. The experiments did not isolate either choice's contribution to search performance.

Decoder awareness enters through the reliability objective, which ranks codes by syndrome-class probabilities across all syndromes; distance remains a feasibility requirement. Distance--multiplicity instead summarizes minimum-weight logical operators. The signals can rank codes differently; their relative value for finite search is empirical. Our methodological contribution connects design objectives, bounded evaluation, and selection; candidate generation uses an existing annealing framework.

Under valid bounds, an upper endpoint below the requirement certifies feasibility. A minimum-cost claim additionally requires excluding every cheaper menu candidate. The \((40,2)\) example meets both conditions (Sec.~\ref{b.-selecting-codes-with-a-failure-requirement}), establishing minimum cost within that menu.

Interval refinement resolves uncertain comparisons among known candidates; pool expansion can reveal better choices, as the union menu illustrates (Sec.~\ref{v.-results}). Search-time hypervolume compares outcomes in truncated reliability coordinates, while refined bounds support candidate selection.

\subsection{Where the reliability signal helps}\label{c.-evidence-for-the-reliability-signal-and-its-limits}

The common-kernel comparison supports decoder-aware guidance for the tested design task, but its benefit depends on the comparator and cell. The distance comparison is consistently positive; the distance--multiplicity comparison includes nonsignificant and negative differences. This variation helps identify where a decoder-defined reliability signal is informative for finite search.

The results are not uniform. At \((40,4)\), the mean difference from distance--multiplicity is negative and remains negative in the doubled-budget campaign. At \((14,4)\), the comparison with distance partly reflects feasibility: 30 of 30 decoder-aware runs and 13 of 30 distance runs reach a feasible design. Application hypervolume measures both access to feasible designs and the quality of the resulting menus. At \(p=0.03\), the \((20,2)\) difference from distance--multiplicity is slightly negative and nonsignificant; the penalty scan covers only two cells (Secs.~\ref{d.-search-budget-dependence-and-the-union-menus}--\ref{e.-penalty-and-physical-error-sensitivity} and Appendix~\ref{app:sensitivity}).

The three primary arms share a search kernel, evaluator, paired seeds, and query budgets, enabling a controlled comparison of their guidance signals. Normalization affects annealing energies and trajectories; lower-cost proxies may remain useful in other search regimes. The single-objective baselines also change initialization and penalty treatment and consequently support a workflow-level comparison. Statistical inference concerns seeded runs under the sign-flip assumptions of Sec.~\ref{e.-paired-inference-and-multiplicity-correction}. The joint Holm correction, exploratory sensitivity tests, and tuning-seed overlap are specified in Sec.~\ref{iv.-search-and-evaluation-protocol} and Appendix~\ref{app:search-execution}.

\subsection{Limitations and next evaluations}\label{d.-noise-dependent-design-variables-and-the-next-evaluation-requirements}

Measurement noise changes what must be selected: a code and a measured check set jointly determine the available reliability--cost choices. Within this fixed main pool, the joint rule searches a superset of the code-first options, so its weak advantage in the truncated objective follows by construction. The empirical results quantify the gap, identify where it vanishes, and test whether intervals separate the selected designs. The weak-gain \((14,4)\) case in Fig.~\ref{fig:noisebudget}, with equal endpoints at nine of 19 budgets for \(q=p\), shows that the benefit depends on the menu and noise setting. The ranking changes reported in Sec.~\ref{b.-code-ordering-and-the-effect-of-redundancy} likewise show that code-capacity ordering need not persist. 

The main limits of the design conclusions are:
\begin{itemize}
\item \emph{Noise and decoding:} independent data and measurement errors, perfect extraction or strict single-round noisy extraction (residual syndrome errors count as failures, without additional ideal recovery), and sector-wise DML; no repeated extraction, correlated faults, or circuit-level fault propagation. Single-shot fault-tolerance guarantees require further structural and noise-control properties \cite{bombin2015}.
\item \emph{Cost model:} \(B_*\) and \(B_{\mathrm{meas}}\) count data-qubit participations but omit routing, ancillas, and scheduling. Flag-based extraction illustrates how circuit choices affect ancilla use and propagated faults \cite{chao2018}.
\item \emph{Pool and decoder:} noisy-syndrome choices are restricted to the fixed perfect-syndrome pool and tested redundancy grid. Decoder optimization and transfer of these rankings to other decoders require separate evaluation; belief propagation with ordered-statistics post-processing is one relevant comparison \cite{roffe2020}.
\item \emph{Statistical scope:} search inferences concern seeded runs under the stated sign-flip assumption, joint Holm correction, and tuning-seed overlap, not a sample of all CSS codes.
\end{itemize}
Applying the rankings to a device requires evaluating its extraction procedure, cost model, and decoder.

Computational cost also constrains the tested approach. Complexity barriers for optimal decoding with degeneracy are known for general stabilizer codes \cite{iyer2015}, without by themselves classifying the restricted CSS family studied here. Truncation reduces error-pattern enumeration to \(\sum_{w=0}^{W}\binom nw\), but the implemented minimum-basis calculation remains exponential in stabilizer rank, and other evaluation steps add cost. The experiments evaluate codes with up to 40 physical qubits. Measuring end-to-end costs would clarify practical efficiency. Tighter numerical enclosures and broader searches address the two distinct decision limits identified above: uncertainty about known candidates and the possibility of missing better ones.

\section{Conclusion}\label{viii.-conclusion}

We have formulated CSS code design at fixed encoding resources as a decoder-aware multiobjective problem, separating sector-wise DML failure from basis-invariant check weight. The resulting candidate menus support choices under a reliability requirement or a check-weight budget. Truncation bounds make these decisions explicit: an upper bound can establish threshold attainment, and lower bounds can exclude cheaper alternatives. The minimum-cost examples establish minimum cost within their evaluated menus. 

The search experiments support decoder-based guidance within the tested common kernel at equal query budgets, with mixed results against distance--multiplicity. Under the reported joint Holm correction on common truncated reliability coordinates, mean final application hypervolume improves significantly over distance in all eight main cells and over distance--multiplicity in five; three comparisons with the latter are nonsignificant, including a negative mean difference at \((40,4)\).  Re-evaluation of the retained codes under strict single-round measurement noise identifies redundancy allocation as an additional design choice: joint selection yields endpoint gains and interval-separated selected-design comparisons at many budgets, while others show no endpoint gain. For a designer using the stated DML and noise model, the procedure is to evaluate a finite pool with failure bounds and choose the lightest code whose upper bound meets the reliability requirement, checking the lower bounds of cheaper candidates for a within-menu minimum-cost claim.

\appendix
\section{Basis-invariant coordinate
components}\label{appendix-a.-basis-invariant-coordinate-components}

This appendix proves the decomposition statement used in Sec.~\ref{c.-distance-and-component-constraints}. All
spaces are binary coordinate spaces on the fixed physical-qubit set
\(E=\{1,\ldots,n\}\). For \(B\subseteq E\), let \(\pi_B\) retain the
coordinates in \(B\) and set the others to zero; \(\mathbb F_2^B\)
denotes its image. For partitions \(\mathcal P\) and \(\mathcal Q\),
write \(\mathcal P\preceq\mathcal Q\) when every block of \(\mathcal P\)
is contained in a block of \(\mathcal Q\). Their join
\(\mathcal P\vee\mathcal Q\) is the finest common coarsening: its blocks
are the equivalence classes generated by membership in a common block of
either partition.

\subsection{A single stabilizer
subspace}\label{a.-a-single-stabilizer-subspace}

Let \(S\subseteq\mathbb F_2^n\) have dimension \(r\), and let \(R\) be a
reduced row-echelon basis matrix, with row vectors \(v_1,\ldots,v_r\)
and pivot coordinates \(p_1,\ldots,p_r\). Form an undirected graph
\(G(S)\) with all \(n\) coordinates as vertices. For each nonpivot
coordinate \(j\) with \(R_{ij}=1\), add the edge \(\{p_i,j\}\). Isolated
vertices are retained. Denote its connected-component partition by
\(\mathcal P_S\).

\textbf{Lemma A1 (coordinate decomposition).} For any partition
\(\mathcal Q\) of \(E\),

\begin{equation}
\begin{gathered}
S=\bigoplus_{B\in\mathcal Q}(S\cap\mathbb F_2^B)
\quad\Longleftrightarrow\quad
\mathcal P_S\preceq\mathcal Q.
\end{gathered}
\label{eq:v9-43}
\end{equation}

Consequently, \(\mathcal P_S\) is the unique finest coordinate
decomposition of \(S\).

\emph{Proof.} A coordinate decomposition over \(\mathcal Q\) is
equivalent to \(\pi_B(S)\subseteq S\) for every block \(B\):
decomposition makes each projection a summand, and conversely the
projections sum to each vector with disjoint supports.

Suppose first that \(S\) decomposes over \(\mathcal Q\), and fix a row
\(v_i\). Let \(B_i\) be the block containing its pivot \(p_i\). Then
\(\pi_{B_i}v_i\in S\). The pivot coordinates of \(v_i\) are \(1\) at
\(p_i\) and \(0\) at every other pivot, and this remains true after
projection onto \(B_i\). For any \(v\in S\), its pivot coordinates uniquely determine its expansion in the basis rows:

\begin{equation}
\begin{gathered}
v=\sum_{h=1}^{r}v_{p_h}v_h.
\end{gathered}
\label{eq:v9-44}
\end{equation}

Thus \(\pi_{B_i}v_i=v_i\), so the entire support of \(v_i\) lies in
\(B_i\). Every edge of \(G(S)\) therefore lies inside a block of
\(\mathcal Q\), which implies \(\mathcal P_S\preceq\mathcal Q\).

Conversely, each row \(v_i\) is supported inside the graph component
containing \(p_i\): its nonpivot support is joined directly to that
pivot. If \(\mathcal P_S\preceq\mathcal Q\), each basis row lies within
a block of \(\mathcal Q\). Grouping the basis rows by those blocks gives
subspaces with disjoint coordinate supports whose direct sum is \(S\).
These subspaces equal \(S\cap\mathbb F_2^B\), proving the decomposition.
The two implications also show that the decomposition is unique and that \(\mathcal P_S\) is the finest such partition. \(\square\)

The partition agrees with the connected-component partition of the
column matroid represented by \(R\). To see this directly, a nonzero
nonpivot column \(j\) has fundamental circuit

\begin{equation}
\begin{gathered}
\{j\}\cup\{p_i:R_{ij}=1\}.
\end{gathered}
\label{eq:v9-45}
\end{equation}

The pivot columns are independent unit vectors, so this dependency is
minimal. Every graph edge consequently connects two elements of a
circuit. Conversely, after rows and columns are grouped by graph
component, \(R\) is block diagonal apart from zero columns. A circuit
cannot cross two such blocks, since a dependency crossing independent
row blocks splits into smaller dependencies. Thus circuit connectivity
and graph connectivity coincide. A zero column is a loop and remains a
singleton; a pivot with no incident edge is a coloop and also remains a
singleton. When \(r=0\), all columns are loops and the graph has \(n\)
singleton components. This is the fundamental-circuit description of matroid components \cite{oxley}.

\subsection{The two CSS stabilizer
subspaces}\label{b.-the-two-css-stabilizer-subspaces}

\textbf{Proposition A2 (CSS components).} Let \(C=(S_X,S_Z)\) satisfy
the CSS commutation condition \(S_X\perp S_Z\). Its unique finest
coordinate-decomposing partition is

\begin{equation}
\begin{gathered}
\mathcal P_C=\mathcal P_{S_X}\vee\mathcal P_{S_Z}.
\end{gathered}
\label{eq:v9-46}
\end{equation}

Equivalently, it is the connected-component partition of the union of
the two graphs constructed above. It depends only on \(S_X\) and
\(S_Z\), not on the generator matrices chosen to represent them.

\emph{Proof.} By Lemma A1, a partition \(\mathcal Q\) decomposes both
subspaces if and only if

\begin{equation}
\begin{gathered}
\mathcal P_{S_X}\preceq\mathcal Q,
\qquad
\mathcal P_{S_Z}\preceq\mathcal Q.
\end{gathered}
\label{eq:v9-47}
\end{equation}

These conditions are equivalent to \(\mathcal P_C\preceq\mathcal Q\) by
the definition of the join. The join itself therefore decomposes both
spaces and is finer than every common decomposing partition.
Connectivity in the union graph takes exactly the transitive closure
that defines this join. Finally, each single-space partition is
characterized by its coordinate-decomposition property, which depends
only on the subspace. Invertible changes of generators preserve both
subspaces and hence preserve their join. \(\square\)

For each block \(c\in\mathcal P_C\), set

\begin{equation}
\begin{gathered}
n_c=|c|,\\
r_{\alpha,c}=\dim(S_\alpha\cap\mathbb F_2^c),
\\ \alpha\in\{X,Z\}.
\end{gathered}
\label{eq:v9-48}
\end{equation}

The restricted spaces remain orthogonal. Inside \(\mathbb F_2^c\), this
implies \(r_{X,c}+r_{Z,c}\le n_c\), and their CSS factor encodes

\begin{equation}
\begin{gathered}
k_c=n_c-r_{X,c}-r_{Z,c}\ge0.
\end{gathered}
\label{eq:v9-49}
\end{equation}

Because both subspaces decompose as direct sums,

\begin{equation}
\begin{gathered}
\sum_c r_{\alpha,c}=r_\alpha,\\
\sum_c k_c=n-r_X-r_Z=k.
\end{gathered}
\label{eq:v9-50}
\end{equation}

A block with \(k_c=0\) is a stabilizer-state factor. Requiring
\(k_c\ge1\) for every block excludes these factors while allowing direct
sums of components that each encode logical information. The
decomposition is over the fixed physical coordinates; it is not a search
over additional Clifford transformations or over different
code-equivalence notions.

\subsection{Implementation
correspondence}\label{c.-implementation-correspondence}

The component routine first reduces both generator matrices to
reduced row-echelon bases, then joins every nonpivot support coordinate of a row
to its pivot. Connected groups give \(\mathcal P_C\). Each basis row
lies in exactly one resulting group, so counting the \(X\)- and
\(Z\)-basis rows in a group gives \(r_{X,c}\) and \(r_{Z,c}\) without a
further rank computation. Zero columns remain in the vertex set even if
no row meets them.

Once the reduced bases are available, there are at most
\(r_X(n-r_X)+r_Z(n-r_Z)\) candidate pivot--nonpivot edges. The cost of
obtaining the reduced bases is separate from this graph construction.

The coordinate decomposition depends on the stabilizer subspaces, whereas Tanner-graph connectivity can change under generator row operations.

\section{Search execution, validation, and analysis details}\label{app:search-execution}

This appendix gives chain-level search and checkpoint details for Sec.~\ref{iv.-search-and-evaluation-protocol}, the analysis settings, and the independent validation of Sec.~\ref{app:validation-details}.

All main-experiment runs used qecpareto v0.3.3.

\subsection{Chain initialization and adaptation}\label{app:chains}

For a given seed, all arms receive the same initial population of 24
random CSS codes at the prescribed ranks. The primary arms start their
nine chains from the same evenly spaced indices in this population.
Chains are advanced in round-robin order. A proposal with energy change
\(\Delta E\le0\) is accepted; otherwise acceptance follows the
Metropolis rule. Each chain starts at temperature \(0.03\), and its
temperature is adapted every 40 proposals using its observed positive
energy changes, with a target uphill acceptance decreasing geometrically
from \(0.25\) to \(0.01\) over the query budget.

\subsection{Checkpoint extraction}\label{app:checkpoints}

The primary outcome is final per-run HV. We also report feasibility and
mean HV at 25\%, 50\%, 75\%, and 100\% of the query budget. Checkpoints
are recorded every 100 queries and at termination; for a requested
budget fraction, we use the largest recorded HV at or before the
target query count, with zero if no checkpoint is available. Curves
average over all 30 seeds, retaining zero-valued failures.

\subsection{Evaluation and analysis details}\label{app:analysis-details}

Each evaluator call, including initialization, rejected proposals, repeated codes, and cache hits, consumes one query. The main experiment uses 24 initialization evaluations per run. For an uncached code, the evaluator computes the logical weight spectra, distance and multiplicity, minimum-basis cost, truncated failure at \(W=4\), and code components. Equal query counts need not give equal distinct-code counts, uncached evaluations, or wall-clock times.

The 108-run tuning phase fixed settings before the main experiment; seeds 1009 and 1046 also occur among main runs, so those runs are not wholly independent of tuning. The sign-flip Monte Carlo calculation uses random seed 20261015. The frozen protocol specified separate Holm corrections for two 16-comparison families; the reported analysis applies one joint 32-test correction. The separate corrections leave all primary significance classifications unchanged at 0.05, although adjusted p values can differ. Thresholds in the search-time screens for \((14,4)\) and \((20,4)\) were chosen post hoc. In the measurement-noise extension, the second-stage tie order was specified during analysis, not prespecified in the original analysis.
\subsection{Validation details}\label{app:validation-details}
The independent validation set contains 400 codes at each of \(n=10,12,14,16,20\), including feasible and infeasible codes, and is distinct from the 477-code main pool. Across these 2000 codes, full-enumeration DML failure lies within its \(W=4\) interval at absolute tolerance \(10^{-14}\). Figure~\ref{fig:validation}(b) displays \((U_W-L_W)/P_D\) for validation codes; the boxes and whiskers show variation among codes. Final-menu widths instead use \((U_W-L_W)/L_W\). The finite validation checks numerical consistency, whereas Proposition 1 establishes the general truncation guarantee.

\section{Search-budget and parameter sensitivity}\label{app:sensitivity}

The five secondary campaigns each contain 180 runs, totaling 900 runs, and are separate from the 1200-run main experiment and the 108 tuning runs. Their p values are unadjusted and separate from the main 32-test Holm family; the raw Monte Carlo floor is \(1/100001\), shown as 0.000010. The budget comparisons and parameter scans below retain their respective cells and query budgets.

\begin{table*}[t]
\caption{Sensitivity of the decoder-aware minus \((d,m)\)
final-HV difference. Budgets are 20,000 queries except for the
explicitly identified 80,000-query rows. Each variant is a separate
180-run campaign with two cells and three signals.}
\label{tab:r4}
\centering
\begin{tabular}{@{}
  p{(\linewidth - 8\tabcolsep) * \real{0.1667}}
  p{(\linewidth - 8\tabcolsep) * \real{0.1667}}
  p{(\linewidth - 8\tabcolsep) * \real{0.2222}}
  p{(\linewidth - 8\tabcolsep) * \real{0.2222}}
  p{(\linewidth - 8\tabcolsep) * \real{0.2222}}@{}}
\toprule

\begin{minipage}[b]{\linewidth}\raggedright
Variant
\end{minipage} & \begin{minipage}[b]{\linewidth}\raggedright
\((n,k)\)
\end{minipage} & \begin{minipage}[b]{\linewidth}\raggedleft
Mean \(\Delta\mathrm{HV}\)
\end{minipage} & \begin{minipage}[b]{\linewidth}\raggedleft
Positive/negative pairs
\end{minipage} & \begin{minipage}[b]{\linewidth}\raggedleft
Raw p
\end{minipage} \\
\midrule

80,000 queries & (40,2) & +0.002253 & 28/2 & 0.000010 \\
80,000 queries & (40,4) & \ensuremath{-}0.000927 & 10/20 & 0.081019 \\
\(\lambda=0.3\) & (20,2) & +0.001465 & 19/11 & 0.523305 \\
\(\lambda=0.3\) & (30,2) & +0.012042 & 30/0 & 0.000010 \\
\(\lambda=3\) & (20,2) & +0.001465 & 19/11 & 0.523305 \\
\(\lambda=3\) & (30,2) & +0.012042 & 30/0 & 0.000010 \\
\(p=0.001\) & (20,2) & +0.009313 & 23/7 & 0.026850 \\
\(p=0.001\) & (30,2) & +0.003388 & 28/2 & 0.000020 \\
\(p=0.03\) & (20,2) & \ensuremath{-}0.000650 & 13/17 & 0.512255 \\
\(p=0.03\) & (30,2) & +0.011928 & 30/0 & 0.000010 \\
\bottomrule
\end{tabular}
\end{table*}

\subsection{Search-budget dependence and pooled menus}\label{app:budget-results}

Table~\ref{tab:r4} includes the doubled-budget comparisons. The trajectory data do not support a fixed ordering of the three signals
at every checkpoint. For example, at \((14,2)\) the final mean HV of the
\((d,m)\) arm is 0.3655, below the distance arm's 0.3668, although the
decoder-aware arm reaches 0.3715. At \((40,2)\) the \((d,m)\) mean is
higher than the decoder-aware mean at the 25\%, 50\%, and 75\%
checkpoints before the small positive decoder-aware difference at the
final budget. These changes motivate examining how the comparison depends on budget, rather than assuming that the final ordering holds throughout the run.

In the separate 80,000-query campaign, the decoder-aware minus \((d,m)\)
mean difference at \((40,2)\) is +0.002253, with 28 positive and two
negative pairs and raw sign-flip \(p\simeq10^{-5}\). At \((40,4)\) the
difference remains negative, \ensuremath{-}0.000927, with 10 positive and 20 negative
pairs and raw \(p=0.0810\). The larger budget therefore yields a
detectable difference in the former comparison but not the latter. These
are within-campaign comparisons; they do not constitute a separate test
of the change in the between-arm effect across budgets.

The longer campaign also supplies candidates that improve the menus. By
the interval-dominance test, its retained candidates dominate 99 of the
117 main-menu entries at \((40,2)\) and 91 of 130 at \((40,4)\). Merging
the two retained sets, deduplicating their code representations,
and filtering again gives union menus of 111 and 137 candidates,
respectively. A smaller retained count can accompany an improved menu
because new candidates remove several older entries.

The \((40,2)\) union menu extends from minimum cost \(B_*=114\) to a
lowest-lower-bound candidate at \(B_*=301\) with interval approximately
\([9.21395,9.24186]\times10^{-5}\). The measurement-noise analysis in Sec.~\ref{vi.-extension-to-noisy-syndrome-measurements} uses the original
477 candidates.

\subsection{Penalty and physical-error sensitivity}\label{app:parameter-results}

The penalty and physical-error campaigns examine only \((n,k)=(20,2)\)
and \((30,2)\), with three signals and 30 paired seeds per cell. Table~\ref{tab:r4} summarizes comparisons against \((d,m)\). Its p values are
unadjusted; these exploratory
analyses are separate from the 32-test Holm family of the main
experiment.

Changing \(\lambda\) from 1 to 0.3 or 3 leaves final HV unchanged for every paired seed of the decoder-aware and \((d,m)\) arms in the two tested cells; the distance-arm mean changes by at most \(7.25\times10^{-4}\). This limited sensitivity is specific to the tested cells and does not imply equivalence to a hard constraint.

At \(p=0.001\), the decoder-aware signal has positive mean differences
from \((d,m)\) in both tested cells, with raw p values 0.02685 and
approximately \(2\times10^{-5}\). At \(p=0.03\), the \((20,2)\)
difference is slightly negative and not detected by the test, while
\((30,2)\) retains a positive difference. The decoder-aware signal has
higher mean HV than distance in all ten cell--variant comparisons, with
raw \(p<10^{-3}\). Since the decoder objective and its HV normalization
change with \(p\), these comparisons are interpreted within each
physical-error setting.

\begin{table*}[t]
\caption{Basis-only measurement noise at \(p=q=0.01\). The
nondominance columns use \((B_*,\widehat P_{\rm ds})\) and do not
certify true Pareto membership. Correlations are descriptive Spearman
coefficients computed with average ranks for ties, between the
code-capacity lower endpoint and \(\widehat P_{\rm ds}\).}
\label{tab:m1}
\centering
\begin{tabular}{@{}
  p{(\linewidth - 8\tabcolsep) * \real{0.1579}}
  p{(\linewidth - 8\tabcolsep) * \real{0.2105}}
  p{(\linewidth - 8\tabcolsep) * \real{0.2105}}
  p{(\linewidth - 8\tabcolsep) * \real{0.2105}}
  p{(\linewidth - 8\tabcolsep) * \real{0.2105}}@{}}
\toprule

\begin{minipage}[b]{\linewidth}\raggedright
\((n,k)\)
\end{minipage} & \begin{minipage}[b]{\linewidth}\raggedleft
Input codes
\end{minipage} & \begin{minipage}[b]{\linewidth}\raggedleft
Distinct nondominated coordinates
\end{minipage} & \begin{minipage}[b]{\linewidth}\raggedleft
Nondominated code records
\end{minipage} & \begin{minipage}[b]{\linewidth}\raggedleft
Spearman \(\rho\), \(t=0\)
\end{minipage} \\
\midrule

(14,2) & 13 & 6 & 8 & 0.906 \\
(14,4) & 4 & 4 & 4 & 1.000 \\
(20,2) & 36 & 13 & 13 & 0.951 \\
(20,4) & 40 & 24 & 24 & 0.969 \\
(30,2) & 61 & 22 & 22 & 0.970 \\
(30,4) & 76 & 24 & 24 & 0.960 \\
(40,2) & 117 & 39 & 39 & 0.971 \\
(40,4) & 130 & 31 & 31 & 0.975 \\
\bottomrule
\end{tabular}
\end{table*}

\section{Additional summaries for noisy syndrome measurements}\label{app:noise-summaries}

The results below use the fixed 477-code main pool, the strict single-round model of Sec.~\ref{a.-model-measured-checks-and-evaluation-bounds}, \(p=0.01\), and \(q\in\{0.01,0.005\}\). The measured rows, redundancy grid, and selection rules are defined in Sec.~\ref{vi.-extension-to-noisy-syndrome-measurements}. Truncated values are lower endpoints, subject to the numerical qualification stated there.

\subsection{Implementation consistency checks}\label{app:noise-checks}

We evaluated 5,644 noisy-syndrome cases across the 477 codes, both
values of \(q\), and the redundancy grids. An additional 477 evaluations
at \(q=0\) recover the code-capacity lower endpoints with a maximum
relative difference of \(3.17\times10^{-13}\); all basis-weight checks
agree. These checks assess implementation consistency and do not add
independent code samples.
For tied check weights, row selection uses the integer row label as a deterministic tie break.

\subsection{Code ordering and redundancy summaries}\label{app:noise-ordering}

At \(q=p\), the cell medians of basis-only \(\widehat P_{\rm ds}\) range
from 0.1068 to 0.1744. The median within-code ratio to the
code-capacity lower endpoint ranges from 8.62 to 491.36 across cells.
Redundancy reduces the truncated values: the median within-code ratio
for one additional row per sector relative to basis-only measurement
lies between 0.688 and 0.891; at \(t_{\max}\), it lies between 0.145 and
0.361. These ratios compare evaluation endpoints and summarize codes
within each cell.

Table~\ref{tab:m1} describes the basis-only menus using the truncated values. It
separates distinct objective coordinates from code records because
equal-coordinate codes need not be the same code.

The strong basis-only correlations coexist with changes in the available
trade-offs. In the six cells with \(n\ge20\), only 23.8\%--60.0\% of the
code records remain nondominated under this point-value comparison. All
four records remain at \((14,4)\). At maximum tested redundancy, rank
correlations are 0.808--0.933 for \(n\ge20\), but fall to 0.623 and
0.400 in the two 14-qubit cells. Thus the data show substantial association in several cells, but the exact code ordering is not preserved under measurement noise.

These ordering statistics are descriptive; interval comparisons between selected designs are reported in Sec.~\ref{d.-what-the-intervals-establish}.

\section*{Code and data availability}
The code and data
supporting this study are not publicly available at the time of this
preprint. We plan to make them publicly available following submission
to a journal or publication of the peer-reviewed article, with the code
released as open-source software.

\section*{ACKNOWLEDGMENT}
This study was financed, in part, by the São Paulo Research Foundation (FAPESP), Brasil. Process Numbers \#2025/25586-5 (postdoctoral fellowship to Y.C.) and \#2021/02120-0. Generative AI tools (OpenAI ChatGPT and Anthropic Claude) were used to assist with software development, and ChatGPT was used for language editing. The authors reviewed and verified all AI-assisted output and take full responsibility for the manuscript and the software.

\onecolumngrid\twocolumngrid
\bibliography{main}

\begin{thebibliography}{27}
\bibitem{terhal2015} \href{https://doi.org/10.1103/RevModPhys.87.307}{B. M. Terhal, Quantum error correction for quantum memories, Rev. Mod. Phys. \textbf{87}, 307 (2015)}.

\bibitem{google2025} \href{https://doi.org/10.1038/s41586-024-08449-y}{Google Quantum AI and Collaborators, Quantum error correction below the surface code threshold, Nature \textbf{638}, 920 (2025)}.

\bibitem{bluvstein2026} \href{https://doi.org/10.1038/s41586-025-09848-5}{D. Bluvstein, A. A. Geim, S. H. Li, S. J. Evered, J. P. Bonilla Ataides, G. Baranes, A. Gu, T. Manovitz, M. Xu, M. Kalinowski, et al., A fault-tolerant neutral-atom architecture for universal quantum computation, Nature \textbf{649}, 39 (2026)}.

\bibitem{putterman2025} \href{https://doi.org/10.1038/s41586-025-08642-7}{H. Putterman, K. Noh, C. T. Hann, G. S. MacCabe, S. Aghaeimeibodi, R. N. Patel, M. Lee, W. M. Jones, H. Moradinejad, R. Rodriguez, et al., Hardware-efficient quantum error correction via concatenated bosonic qubits, Nature \textbf{638}, 927 (2025)}.

\bibitem{bravyi2024} \href{https://doi.org/10.1038/s41586-024-07107-7}{S. Bravyi, A. W. Cross, J. M. Gambetta, D. Maslov, P. Rall, and T. J. Yoder, High-threshold and low-overhead fault-tolerant quantum memory, Nature \textbf{627}, 778 (2024)}.

\bibitem{pecorari2025} \href{https://doi.org/10.1038/s41467-025-56255-5}{L. Pecorari, S. Jandura, G. K. Brennen, and G. Pupillo, High-rate quantum LDPC codes for long-range-connected neutral atom registers, Nat. Commun. \textbf{16}, 1111 (2025)}.

\bibitem{scruby2026} \href{https://doi.org/10.1103/67xf-zdjb}{T. R. Scruby, T. Hillmann, and J. Roffe, High-threshold, low-overhead and single-shot decodable fault-tolerant quantum memory, PRX Quantum \textbf{7}, 020310 (2026)}.

\bibitem{zhou2025} \href{https://doi.org/10.1038/s41586-025-09543-5}{H. Zhou, C. Zhao, M. Cain, D. Bluvstein, N. Maskara, C. Duckering, H.-Y. Hu, S.-T. Wang, A. Kubica, and M. D. Lukin, Low-overhead transversal fault tolerance for universal quantum computation, Nature \textbf{646}, 303 (2025)}.

\bibitem{hillmann2025} \href{https://doi.org/10.1038/s41467-025-63214-7}{T. Hillmann, L. Berent, A. O. Quintavalle, J. Eisert, R. Wille, and J. Roffe, Localized statistics decoding for quantum low-density parity-check codes, Nat. Commun. \textbf{16}, 8214 (2025)}.

\bibitem{perlin} \href{https://arxiv.org/abs/2607.19574}{M. A. Perlin, M. Steinberg, and B. Criger, Quantum codes from classical annealing, arXiv:2607.19574 (2026)}.

\bibitem{webster} \href{https://doi.org/10.1109/TQE.2025.3538934}{M. A. Webster and D. E. Browne, Engineering Quantum Error Correction Codes Using Evolutionary Algorithms, IEEE Trans. Quantum Eng. \textbf{6}, 3100514 (2025)}.

\bibitem{he} \href{https://arxiv.org/abs/2502.14372}{A. Y. He and Z.-W. Liu, Discovering highly efficient low-weight quantum error-correcting codes with reinforcement learning, arXiv:2502.14372 (2025)}.

\bibitem{lee} \href{https://arxiv.org/abs/2609.30069}{J. Y. Lee, K. Okada, N. Maskara, K. Kasai, and H. Zhou, Design Principles for Ultra-High-Rate Quantum Codes, arXiv:2609.30069 (2026)}.

\bibitem{nautrup} \href{https://doi.org/10.22331/q-2019-12-16-215}{H. P. Nautrup, N. Delfosse, V. Dunjko, H. J. Briegel, and N. Friis, Optimizing Quantum Error Correction Codes with Reinforcement Learning, Quantum \textbf{3}, 215 (2019)}.

\bibitem{su} \href{https://doi.org/10.1103/PhysRevApplied.23.034048}{V. P. Su, C. Cao, H.-Y. Hu, Y. Yanay, C. Tahan, and B. Swingle, Discovery of optimal quantum codes via reinforcement learning, Phys. Rev. Applied \textbf{23}, 034048 (2025)}.

\bibitem{ashikhmin} \href{https://doi.org/10.1109/JSAC.2020.2968997}{A. Ashikhmin, C.-Y. Lai, and T. A. Brun, Quantum Data-Syndrome Codes, IEEE J. Sel. Areas Commun. \textbf{38}, 449 (2020)}.

\bibitem{calderbank1996} \href{https://doi.org/10.1103/PhysRevA.54.1098}{A. R. Calderbank and P. W. Shor, Good quantum error-correcting codes exist, Phys. Rev. A \textbf{54}, 1098 (1996)}.

\bibitem{steane1996} \href{https://doi.org/10.1103/PhysRevLett.77.793}{A. M. Steane, Error correcting codes in quantum theory, Phys. Rev. Lett. \textbf{77}, 793 (1996)}.

\bibitem{gottesman1997} \href{https://arxiv.org/abs/quant-ph/9705052}{D. Gottesman, \textit{Stabilizer Codes and Quantum Error Correction}, Ph.D. thesis, California Institute of Technology (1997), arXiv:quant-ph/9705052}.

\bibitem{iyer2015} \href{https://doi.org/10.1109/TIT.2015.2422294}{P. Iyer and D. Poulin, Hardness of decoding quantum stabilizer codes, IEEE Trans. Inf. Theory \textbf{61}, 5209 (2015)}.

\bibitem{oxley} \href{https://doi.org/10.1093/acprof:oso/9780198566946.001.0001}{J. Oxley, \textit{Matroid Theory}, 2nd ed. (Oxford University Press, 2011)}.

\bibitem{chao2018} \href{https://doi.org/10.1103/PhysRevLett.121.050502}{R. Chao and B. W. Reichardt, Quantum error correction with only two extra qubits, Phys. Rev. Lett. \textbf{121}, 050502 (2018)}.

\bibitem{zitzler} \href{https://doi.org/10.1109/4235.797969}{E. Zitzler and L. Thiele, Multiobjective evolutionary algorithms: A comparative case study and the strength Pareto approach, IEEE Trans. Evol. Comput. \textbf{3}, 257 (1999)}.

\bibitem{holm} \href{https://www.jstor.org/stable/4615733}{S. Holm, A Simple Sequentially Rejective Multiple Test Procedure, Scand. J. Stat. \textbf{6}, 65 (1979)}.

\bibitem{dennis2002} \href{https://doi.org/10.1063/1.1499754}{E. Dennis, A. Kitaev, A. Landahl, and J. Preskill, Topological quantum memory, J. Math. Phys. \textbf{43}, 4452 (2002)}.

\bibitem{bombin2015} \href{https://doi.org/10.1103/PhysRevX.5.031043}{H. Bombín, Single-shot fault-tolerant quantum error correction, Phys. Rev. X \textbf{5}, 031043 (2015)}.

\bibitem{roffe2020} \href{https://doi.org/10.1103/PhysRevResearch.2.043423}{J. Roffe, D. R. White, S. Burton, and E. T. Campbell, Decoding across the quantum low-density parity-check code landscape, Phys. Rev. Research \textbf{2}, 043423 (2020)}.
\end{thebibliography}
\end{document}